%% file: main.tex
\PassOptionsToPackage{table}{xcolor}  
\documentclass[sigconf]{acmart}

\usepackage{caption}
\usepackage{subcaption}
\usepackage{graphicx}
\usepackage[inline]{enumitem}

\usepackage{listings}

\AtBeginDocument{%
  }

\usepackage[most]{tcolorbox}
\usepackage{fontawesome5}
\newtcolorbox{greentextbox}[1][]{%
    colback=green!10,
    colframe=green!15,
    notitle,
    rounded corners,
    enhanced,
    breakable,
    left=0pt,
    right=0pt,
    top=0pt,
    bottom=0pt
    }

\newtcolorbox{redtextbox}[1][]{%
    colback=red!10,
    colframe=red!15,
    notitle,
    rounded corners,
    enhanced,
    breakable,
    left=0pt,
    right=0pt,
    top=0pt,
    bottom=0pt
    }

\newtcolorbox{bluetextbox}[1][]{%
    colback=blue!10,
    colframe=blue!15,
    notitle,
    rounded corners,
    enhanced,
    breakable,
    left=0pt,
    right=0pt,
    top=0pt,
    bottom=0pt
    }

\setcopyright{acmlicensed}
\copyrightyear{2018}
\acmYear{2018}
\acmDOI{XXXXXXX.XXXXXXX}
\acmConference[Conference acronym 'XX]{Make sure to enter the correct
  conference title from your rights confirmation email}{June 03--05,
  2018}{Woodstock, NY}
\acmISBN{978-1-4503-XXXX-X/2018/06}

\definecolor{rgb:144,238,144}{RGB}{144,238,144}
\definecolor{rgb:194,230,153}{RGB}{194,230,153}
\definecolor{rgb:241,224,168}{RGB}{241,224,168}
\definecolor{rgb:245,183,177}{RGB}{245,183,177}
\definecolor{rgb:255,160,160}{RGB}{255,160,160}

\begin{document}

\title{The Hidden Threat in Plain Text:\\Attacking RAG Data Loaders}

\author{Alberto Castagnaro}
\affiliation{%
  \institution{University of Padua}
  \city{Padova}
  \country{Italy}}
\email{alberto.castagnaro@unipd.it}

\author{Umberto Salviati}
\affiliation{%
  \institution{University of Padua}
  \city{Padova}
  \country{Italy}}
\email{umberto.salviati@studenti.unipd.it}

\author{Mauro Conti}
\affiliation{%
  \institution{University of Padua}
  \city{Padova}
  \country{Italy}}
\email{mauro.conti@unipd.it}

\author{Luca Pajola}
\affiliation{%
  \institution{Spritz Matter}
  \city{Padova}
  \country{Italy}}
\email{luca.pajola@spritzmatter.com}

\author{Simeone Pizzi}
\email{sime.pizzi@gmail.com}

\renewcommand{\shortauthors}{Anonymous et al.}
\newcommand{\todo}[1]{\textcolor{red}{TODO: #1}}
\begin{abstract}
Large Language Models (LLMs) have transformed human--machine interaction since ChatGPT's 2022 debut, with Retrieval-Augmented Generation (RAG) emerging as a key framework that enhances LLM outputs by integrating external knowledge. However, RAG's reliance on ingesting external documents introduces new vulnerabilities.
This paper exposes a critical security gap at the data loading stage, where malicious actors can stealthily corrupt RAG pipelines by exploiting document ingestion. 
\par
We propose a taxonomy of 9 knowledge-based poisoning attacks and introduce two novel threat vectors---\textit{Content Obfuscation} and \textit{Content Injection}---targeting common formats (DOCX, HTML, PDF).
Using an automated toolkit implementing 19 stealthy injection techniques, we test five popular data loaders, finding a 74.4\% attack success rate across 357 scenarios. We further validate these threats on six end-to-end RAG systems---including white-box pipelines and black-box services like NotebookLM and OpenAI Assistants---demonstrating high success rates and critical vulnerabilities that bypass filters and silently compromise output integrity.
Our results emphasize the urgent need to secure the document ingestion process in RAG systems against covert content manipulations.

\end{abstract}



\received{20 February 2007}
\received[revised]{12 March 2009}
\received[accepted]{5 June 2009}

\maketitle

%

\input{Sections/01-Introduction}

\input{Sections/02-background}
\input{Sections/03-ThreatLandscape-v2}

\input{Sections/99-threat-model}
\input{Sections/04-PhantomRAG}

\input{Sections/99-resultsPhantom}
\input{Sections/10-Discussions}

\bibliographystyle{ACM-Reference-Format}
\bibliography{bibliography}

\appendix
\input{Sections/Appendix}

\end{document}

%% file: Sections/01-Introduction.tex
\section{Introduction}





  
  
  
  


Large Language Models (LLMs), despite their capabilities~\cite{brown2020language}, suffer from limitations such as hallucinations, outdated knowledge, and the inability to access real-time or proprietary information~\cite{roller2020recipes}, which Retrieval-Augmented Generation (RAG) addresses by integrating external knowledge sources~\cite{lewis2020retrieval}. By retrieving relevant documents from structured or unstructured data and integrating them into the generation, RAG improves the reliability and applicability of answers, making it particularly valuable for enterprise documentation, legal research, and cybersecurity operations.
\par
While LLMs exhibit remarkable capabilities in various domains, they also pose significant and novel security challenges. Their reliance on massive datasets and complex architectures makes them vulnerable to a variety of attacks that can compromise their reliability and security. 
Prompt injection attacks can override system instructions and manipulate model behavior, resulting in unintended
outcomes~\cite{perez2022ignore}. In addition, LLMs can generate biased or harmful content from training data~\cite{weidinger2022taxonomy} and are vulnerable to attacks like model inversion and adversarial examples that extract or manipulate sensitive information~\cite{carlini2021extracting}.
More recently, a Common Vulnerabilities and Exposures (CVE) has been disclosed in Meta's Llama framework exposed AI applications to remote execution attacks (CVE-2024-50050).\footnote{\url{https://nvd.nist.gov/vuln/detail/CVE-2024-50050}}
\par
\paragraph{Contributions.}
Since RAG systems are becoming more and more popular, it is very important to understand how well they are protected against cyberattacks. For this reason, in this paper we propose a study of such systems from a cybersecurity perspective, with the three key contributions:
\begin{enumerate*}
    \item we propose a taxonomy of knowledge-based poisoning attacks against RAG;
    \item we introduce a set of practical techniques for poisoning the knowledge base during data loading;
    \item we evaluate both popular open-source data loading libraries and black-box RAG systems from popular providers, finding that most of them are vulnerable to simple techniques. In this regard, we develop and release PhantomText toolkit, a framework for the automatic generation of poisoned documents.
\end{enumerate*}

%% file: Sections/02-background.tex
\section{Background}\label{sec:background}
\subsection{Retrieval-Augmented Generation}
\subsubsection{Overview}
Retrieval-Augmented Generation (RAG) is a framework that enhances large language models (LLMs) by integrating external knowledge retrieval, improving the relevance and accuracy of outputs~\cite{lewis2020retrieval}. Unlike traditional models, RAG accesses real-time information, making it effective in dynamic contexts.
\par
A RAG pipeline integrates information retrieval with generative AI to improve response accuracy and relevance. It begins with \textit{database generation}, where structured or unstructured data (e.g., documents, PDFs, web pages) are embedded and stored in a \textit{vector database}\footnote{A vector database stores high-dimensional embeddings to enable fast semantic similarity searches.}. Upon receiving a \textit{user query}, relevant documents are retrieved and the results are passed to a \textit{language model}, which incorporates this context to produce accurate, informed responses.
\subsubsection{Building a vector database}
As our research focuses on the security of the \textit{parsing stage} in RAG pipelines, it is cucial to examine how the vector database is built. The reliability of retrieval depends on how data is \textit{preprocessed, tokenized, embedded}, and stored, as flaws in these steps may enable poisoning attacks, adversarial inputs, or inconsistent retrieval.
A RAG pipeline is constructed by
\begin{enumerate*}
    \item gathering relevant data;
    \item loading the data, segmenting it into chunks;
    \item converting the data into \emph{vector embeddings};
    \item storing the data in a vector database for efficient retrieval;
    \item retrieving the relevant chunks at query time and inserting them into the LLM context, to enhance its responses
\end{enumerate*}

\subsection{LLM Jailbreaking}
LLM jailbreaking~\cite{wei2023jailbroken} refers to the act of bypassing or manipulating the safeguards built into a language model, which can lead to unauthorized data access and the generation of harmful content. RAG systems can be vectors for such attacks~\cite{pandora}.

%% file: Sections/03-ThreatLandscape-v2.tex
\section{Taxonomy of Knowledge Base Poisoning Attacks in RAG Systems}\label{sec:threat-landscape}

With the growing adoption of AI systems, security researchers have started studying the security risks of RAG pipelines. To fully understand these risks, we highlight and categorize the goals of potential attackers. We define four main attack families: system integrity attacks, output manipulation attacks, knowledge poisoning attacks, and security and privacy attacks, based on the adversary's objective. To ensure a systematic analysis, we map each attack to the CIA triad which has the advantage of being a familiar framework for most security researchers.

\subsection{RAG and the CIA triad}
RAG integrates information retrieval with generative models to improve content accuracy and relevance, yet its often-overlooked vulnerability to cyberattacks can be systematically analyzed using the CIA triad (Confidentiality, Integrity, and Availability).
\par
Integrity concerns stem from the risk of retrieving or generating inaccurate, tampered with, or adversarial content. Compromised or unreliable sources can lead to misinformation or biased outputs. Availability is equally critical, as RAG systems depend on external databases and APIs, making them vulnerable to denial-of-service attacks or retrieval failures. Securing RAG applications within the CIA framework requires robust access control, data validation, and resilience against adversarial input.
Table~\ref{tab:cia_mapping} shows the mapping between the proposed attacks and the CIA triad.


\begin{table}[ht]
\footnotesize
\centering
\caption{CIA Mapping for RAG Knowledge Base Poisoning Attacks}
\begin{tabular}{|l|c|c|c|}
\toprule
\textbf{Attack} & \textbf{(C)} & \textbf{(I)} & \textbf{(A)} \\ \midrule
\multicolumn{4}{|c|}{\textit{System Integrity Attacks}} \\ \midrule
Pipeline Failure                 &                       &                  & \faCheck                  \\ \hline
Reasoning Overload &                       &                  & \faCheck                  \\ \midrule
\multicolumn{4}{|c|}{\textit{Response Manipulation Attacks}} \\ \midrule
Unreadable Output                &                       &                  & \faCheck                  \\ \hline
Empty Statement Response         &                       &                  & \faCheck                  \\ \hline
Vague Output                 &                       & \faCheck                 & \faCheck                  \\ \midrule
\multicolumn{4}{|c|}{\textit{Knowledge Manipulation Attacks}} \\ \midrule
Bias Injection       &                       & \faCheck                 &                   \\ \hline
Factual Distortion               &                       & \faCheck                 &                   \\ \hline
Outdated Knowledge               &                       & \faCheck                 &                   \\ \midrule
\multicolumn{4}{|c|}{\textit{Security and Privacy Attacks}} \\ \midrule
Sensitive Data Disclosure        & \faCheck                      &                  &                   \\ \bottomrule
\end{tabular}
\label{tab:cia_mapping}
\end{table}

\subsection{System Integrity Attacks}
Attacks that disrupt RAG stability, performance, or efficiency.
\subsubsection{Pipeline Failure (\textbf{A})}
Injected adversarial data causes unexpected software failures, including crashes, infinite loops, or unhandled exceptions.
It affects system availability, as the system may crash or enter an unstable state, making it inaccessible to users.
We did not find references to RAG pipeline failures in the literature, however, Xiao et al.~\cite{xiao2018security} have already discussed DoS attacks in traditional machine learning libraries. 
\begin{greentextbox}
{\scriptsize \faBomb } \textbf{\texttt{A1}: Pipeline Failure.} \label{a1_attack}
Crashes or destabilizes the RAG system through adversarial inputs that trigger software failures, leading to a denial-of-service.
\end{greentextbox}

\subsubsection{Reasoning Overload (\textbf{A, I})}
A recent trend in Large Language Models is \textit{reasoning models}, which consists in models that are explicitly trained to describe how they got to the result before producing a final answer to the user query, thus improving their performance across a number of tasks~\cite{wei2022chain, huang2022towards}.

An LLM Denial-of-Service (DoS) attack can leverage this ``reasoning'' by poisoning the vector database to inject malicious content that forces the LLM to spend excessive computational resources, thereby slowing down inference and making the system less responsive, or even unresponsive, under certain conditions. 
The attack could impact multiple aspects:
\begin{itemize}
    \item \textit{Performance Slowdown \textbf{(A)}}: The attack can result in a significant increase in processing time, severely degrading the user experience.
    It impacts the Availability as the system remains functional but becomes inefficient, reducing usability.
    \item \textit{Excessive Token Production \textbf{(A, I)}}: The inclusion of decoy reasoning problems increases the token count, leading to a higher computational load. This amplifies the cost of operating the system, as each inference requires more tokens to be processed, which can negatively impact the financial sustainability of third-party applications that rely on reasoning LLMs. It affects availability, as an increased computational burden can slow down the system or cause request failures. It affects the integrity, as the attack manipulates the system's expected behavior by forcing unnecessary reasoning steps, altering the intended flow of information processing.
    \item \textit{Resource Strain and Service Denial \textbf{(A)}}: Over time, if multiple users are subjected to similar poisoned queries, the system could become unresponsive, or the service could fail to return a timely response, essentially causing a DoS effect.
\end{itemize}

This attack aligns with the ``OVERTHINK'' attack~\cite{kumar2025overthink}, where adversaries inject decoy reasoning problems to increase token processing during inference, causing the model to ``overthink'' while generating responses.

\begin{greentextbox}
{\scriptsize \faBomb } \textbf{\texttt{A2}: Reasoning Overload.} 
Injecting adversarial content designed to expand excessively upon retrieval, causing increased inference time and resource exhaustion.
\end{greentextbox}


\subsection{Output Manipulation Attacks}
Attacks that reduce response quality, clarity, or usability.

\subsubsection{Unreadable Output (\textbf{A})}
The system may produce nonsensical, illegible, or unusable outputs due to tokenization errors, corrupted embeddings, or external tampering. In some cases, adversarial perturbations can subtly modify inputs to exploit these vulnerabilities. For instance, attackers may inject misleading Unicode characters or embed Base64-encoded segments, formats that are technically valid but not interpretable by the model, causing garbled text, random character sequences, misinterpreted tokens, or otherwise irrelevant output.
These manipulations degrade output interpretability without crashing the system, leaving it operational yet functionally unusable.


\begin{greentextbox}
{\scriptsize \faBomb } \textbf{\texttt{A3}: Unreadable Output.} 
Injecting adversarial text containing excessive special characters or encoding artifacts, leading to responses filled with unreadable symbols.
\end{greentextbox}

\subsubsection{Empty Statement Response (\textbf{A})}
The attack targets output generation, causing the system to produce grammatically correct but meaningless responses that lack actionable context, often due to prompt filters, training-data biases, or safety constraints.
For example, adversarial input might manipulate an RAG system to output generic fallback statements like ``I’m sorry, I cannot provide that information,'' even when the query is legitimate and should have resulted in a specific response. Attackers can exploit this vulnerability by feeding misleading or irrelevant data into the vector database and forcing the system into evasive or empty responses, undermining its utility in critical contexts.
This attack affects availability, as the system responds but provides no meaningful value, disrupting usability.

\begin{greentextbox}
{\scriptsize \faBomb } \textbf{\texttt{A4}: Empty Statement Response.} 
 Causing the system to respond with generic statements such as \textit{``I cannot answer this question.''}
\end{greentextbox}

\subsubsection{Vague Output (\textbf{A, I})}
The system generates vague or misleading responses that lack clear factual grounding.
These responses often lack the necessary specificity, context, or detail needed to effectively address the user's query.  
For example, a RAG-powered assistant might respond with ``There are many factors to consider,'' when asked for a recommendation or solution, leaving the user unsure of what factors are involved or how they should proceed. Such responses often result from adversarial interference that forces the model to retrieve or generate generalized, context-agnostic text, which could be interpreted in numerous ways. This ambiguity undermines the system's reliability and hinders users from making informed decisions, especially in environments where precision is key.
This attack effect is twofold: \textit{(i)} availability, as the model responds but does not deliver actionable insights, making the response functionally useless; (ii) integrity, as the system’s correctness is affected, as it no longer provides clear and accurate information.

\begin{greentextbox}
{\scriptsize \faBomb } \textbf{\texttt{A5}: Vague Output.} 
Poisoning the knowledge base with subtly inconsistent data results in outputs like \textit{``There is no definitive answer to this question.''}
\end{greentextbox}

\subsection{Knowledge Manipulation Attacks}
Attacks altering knowledge, affecting accuracy, bias or timeliness.

\subsubsection{Bias Injection (\textbf{I})}
The knowledge base is poisoned with biased content, which reinforces existing perspectives while reducing access to diverse viewpoints. This causes the RAG system to present skewed, unfair, or ideologically extreme narratives, affecting system integrity, as the responses become unreliable due to systemic bias in the retrieved information.

\begin{greentextbox}
{\scriptsize \faBomb } \textbf{\texttt{A6}: Bias Injection.} 
Poisoning the RAG system with misleading information leading to biased responses favoring one perspective.
\end{greentextbox}

\subsubsection{Factual Distortion (\textbf{I})}
The attack occurs when the system generates or retrieves information that is factually incorrect, misleading, or unverifiable, thus compromising the system’s credibility and reliability.
The attacks affect system integrity, as factual correctness is no longer guaranteed, corrupting the system's reliability.
Factual distortion can be linked to the more general phenomenon of misinformation campaigns (fake news)~\cite{lazer2018science}. 

\begin{greentextbox}
{\scriptsize \faBomb } \textbf{\texttt{A7}: Factual Distortion.} 
Modifying the knowledge base with factually incorrect information, leading the RAG system to propagate misinformation.
\end{greentextbox}

\subsubsection{Outdated Knowledge (\textbf{I})}
This attack obfuscates recent data, causing the RAG system to retrieve outdated information like obsolete studies, rulings, or product specs. Since the system still responds, outdated info may go unnoticed, leading to faulty recommendations or decisions.

\begin{greentextbox}
{\scriptsize \faBomb } \textbf{\texttt{A8}: Outdated Knowledge.} 
Causing the model to base its answers on obsolete knowledge, leading to responses containing inaccurate or outdated information.
\end{greentextbox}

\subsection{Security and Privacy Violations}
Attacks that breach confidentiality or expose sensitive data.

\subsubsection{Sensitive Data Disclosure (\textbf{C})}
Sensitive data disclosure occurs when the system inadvertently reveals confidential information breaching privacy and confidentiality. This can happen when the system retrieves or generates real data that should not be accessible or exposed to users. The disclosed data could include personally identifiable information (PII), such as phone numbers, medical records, financial information, or even sensitive corporate data.
The attack impacts system confidentiality by improperly disclosing sensitive data, causing privacy and security violations.

\begin{greentextbox}
{\scriptsize \faBomb } \textbf{\texttt{A9}: Sensitive Data Disclosure.} 
Making the RAG system disclose sensitive information, such as PII or corporate data, to unauthorized users.
\end{greentextbox}

%% file: Sections/99-threat-model.tex
\section{Threat Model}\label{sec:threat_model}

\paragraph{\textbf{Target}}
The \emph{Target} of the attack is a RAG system (Section~\ref{sec:background}) operated by a business. We assume that before documents are added to the knowledge base of the RAG, they are inspected by one or more employees, using tools such as PDF readers and web browsers to check for abuses. We assume that these employees are not IT experts, but they are experts in the topics covered by the documents, i.e., they are able to recognize documents that contain inaccurate or false information. We call these employees \emph{Inspectors}.

\paragraph{\textbf{Attacker}}
The \emph{Attacker} is a malicious actor that does not have direct access to the \emph{Target}, but is able to inject poisoned documents into the RAG knowledge base by different means. These include, but are not limited to:

\begin{itemize}
    \item \textbf{Supply Chain Attacks}: The attacker could be a service provider, and poison all the documents it provides as documentation to its clients. In this case, any of the clients that imports the documents in a RAG system would become a victim.
    \item \textbf{Web-Based Poisoning}: Attackers manipulate public sources (e.g., wikis, research databases) by injecting false or biased information. A RAG-powered cybersecurity assistant, for example, could retrieve and suggest insecure cryptographic practices, exposing developers to security risks.
    \item \textbf{Insider Threats in Organization}: A malicious employee in a company modifies internal documents used by a RAG system, injecting false compliance guidelines. Employees relying on the system may unknowingly follow misleading legal or financial advice, leading to regulatory violations.
\end{itemize}

\paragraph{\textbf{Goal}}
The attacker seeks to undermine the integrity, reliability, or security of RAG systems by embedding imperceptible manipulations in retrieved knowledge. Their objectives include misleading responses that introduce factual distortions or biases, stealthy information manipulation via obfuscated content, evading detection mechanisms through adversarial encoding, targeted prompt manipulation to exploit hidden triggers, and poisoning the knowledge base to degrade system performance. A detailed breakdown of these goals and their mapping to CIA principles is provided in Section~\ref{sec:threat-landscape}.


To achieve these objectives, the attacker leverages vulnerabilities in the \textit{Data Loading} stage of the RAG pipeline, performing attacks that fall under two broad categories:
\begin{itemize}
    \item \textbf{Content Obfuscation}: Disrupts or distorts existing information in the document using invisible characters, making it harder for models to correctly extract or interpret the original content.
    \item \textbf{Content Injection}: Inserts new, invisible concepts into the document, effectively introducing misleading or fabricated knowledge into the knowledge base.
\end{itemize}

%% file: Sections/04-PhantomRAG.tex
\section{Data Loading Deception Techniques}\label{sec:phantom}
In this section, we describe the techniques we studied to attack the data loading stage of a RAG pipeline.

\subsubsection{Technique Organization}
The attack strategy is organized into two major deception technique families:
\begin{itemize}
    \item \textbf{Content Obfuscation}: This family of attacks aims to disrupt or distort existing information in the document, reducing its readability or interpretability for the AI model. We use techniques such as zero-width characters (e.g., ZWSP, ZWNJ), homoglyph substitutions (e.g., replacing letters with visually similar Unicode characters), and bidirectional (Bidi) reordering to subtly alter the structure of words or sentences without changing their appearance to the human eye.
    \item \textbf{Content Injection}: These attacks introduce new, invisible or hidden concepts into the document, effectively poisoning the knowledge base with misleading or fabricated facts. Techniques include using font size zero to make text visually disappear, positioning text out of the visible margin, and injecting data into document metadata fields (e.g., PDF/XMP metadata), which may be parsed and indexed by AI systems but remain invisible to end users.
\end{itemize}
In our analysis -- as RAG can be set up to handle input documents of different formats -- we examine the structure and formatting capabilities of widely used document types— PDF (~\faFilePdf), HTML  (~\faHtml5), and DOCX (~\faFileWord)—to identify potential threat vectors that enable both content obfuscation and content injection attacks. The rich styling and metadata features of these formats create ideal surfaces for content obfuscation or injection - using invisible characters, style tweaks, embedded objects, and hidden fields - and analyzing their text rendering, metadata handling, and layout exposes unique stealth attack vectors in RAG pipelines.


\subsection{Content Obfuscation}

\subsubsection{Diacritical Marks ~\faFilePdf~\faHtml5}
Diacritical marks are small signs or symbols added to letters to alter their pronunciation, meaning, or distinguish between words in different languages. They include accent marks (e.g., é, ñ), tildes, umlauts, and cedillas, and are commonly used in languages like French, Spanish, and German.
This attack involves combining diacritical marks—Unicode characters that modify a base character—to create visually complex representations of text. 
In our attakcs, we add ten diacritical character to a random letter on the target word.

\subsubsection{Homoglyph Characters~\faFilePdf~\faFileWord~\faHtml5}
This attack replaces characters with visually identical but semantically different Unicode homoglyphs, such as substituting Latin ``A'' (U+0041) with Cyrillic one (U+0410).
Homoglyph substitution attacks have been shown to be effective in attacking traditional ML-based and information-retrieval systems deployed in real-world applications~\cite{grondahl2018all, boucher2022bad, boucher2023boosting}.

\subsubsection{OCR-Poisoning ~\faFilePdf~\faHtml5}
This attack involves introducing subtle perturbations into text parsed by Optical Character Recognition (OCR) systems, particularly in images and complex PDFs. By manipulating visual features that OCR relies on for text extraction, these perturbations can confuse the recognition process, leading to incorrect or misleading interpretations of the original text.
This attack has been demonstrated successful in modern OCR systems~\cite{conti2023turning}.

\subsubsection{Reordering Characters~\faFilePdf~\faFileWord~\faHtml5}
A reordering attack, or bidirectional (Bidi) attack, leverages the Unicode Bidirectional algorithm to reorder characters visually while preserving a different logical encoding order. By strategically embedding right-to-left (RTL) characters within left-to-right (LTR) text, it creates discrepancies between human and machine interpretation. 
Bidi attacks have been demonstrated successful in attacking traditional ML-based and information-retrieval systems deployed in real applications~\cite{boucher2022bad, boucher2023boosting}.

\subsubsection{Zero-Width Characters ~\faFilePdf~\faFileWord~\faHtml5}
This attack leverages zero-width (ZeW) characters—Unicode symbols that take no visible space but alter text representation. Common examples include Zero-Width Space (U+200B), Zero-Width Non-Joiner (U+200C), and Zero-Width Joiner (U+200D). These characters are typically used for text formatting, preventing ligatures, or controlling word breaks in multilingual settings. By inserting them strategically, the attack disrupts tokenization and downstream NLP processing without altering human-readable text.
Zero-width space attacks have been demonstrated successful in attacking traditional machine learning-based and information-retrieval systems deployed in real applications~\cite{pajola2021fall, boucher2022bad, boucher2023boosting}.
We implemented three distinct variants, conducting tests over the 19 distinct ZeW characters:
\begin{enumerate*}
    \item \textit{Mask1 \faFilePdf~\faFileWord~\faHtml5}: Insertion of one ZeW character in the middle of the target word. 
    \item \textit{Mask2 \faFilePdf~\faFileWord~\faHtml5}: Insertion of three ZeW characters in the middle of the target word. 
    \item \textit{Mask3 \faFilePdf~\faFileWord~\faHtml5}: Insertion of many ZeW characters between every character of a target word. 
\end{enumerate*}

\subsection{Content Injection}
\subsubsection{Camouflage Element ~\faFilePdf~\faHtml5}
The technique embeds text behind other elements within a document, effectively rendering it invisible to the user while preserving its presence within the document's structure.
In this work, we inject malicious text behind images.

\subsubsection{Metadata ~\faFilePdf~\faHtml5}
This attack involves adding or altering a document's metadata with malicious or poisoned data that may be inadvertently parsed and processed by document loaders or AI models. We inject random or adversarial elements into the document's metadata (i.e. PDF metadata or HTML \texttt{<head>}). 

\subsubsection{Out-of-Bound Text ~\faFilePdf~\faFileWord~\faHtml5}
This attack involves injecting text outside the visible area of the document, such as beyond the margins or in hidden portions of the webpage or document. While the text becomes invisible to the user because it is rendered outside the viewport, it is still processed by machines. 
We designed two variants:  
\begin{enumerate*}
    \item \textit{Random position ~\faFilePdf~\faHtml5}: The text is placed at random coordinates outside the visible area of the document or webpage, such as the top-left or bottom-right corners.
    \item \textit{Right margin ~\faFileWord}: The text is aligned to the far right, partially pushed off-screen, leaving only a fragment visible.
\end{enumerate*}
\subsubsection{Transparent Text ~\faFilePdf~\faFileWord~\faHtml5}
This attack makes adversarial text the same color as the background, making it invisible to humans while still being processed by AI models. It exploits formatting features in different file types (e.g., HTML, DOCX, PDF).
\begin{enumerate}
    \item \textit{Font color matching the background ~\faFilePdf~\faFileWord~\faHtml5}, with background set to white in the Word format case. 
    \item \textit{Opacity 0 ~\faFilePdf~\faHtml5}: In HTML, setting `\texttt{opacity: 0;}' in CSS (`\texttt{<span style=``opacity: 0;''>hidden text</span>}') makes text invisible to users while still readable by screen readers and AI models. In PDFs, this effect is replicated using transparent text layers or annotations with zero opacity.
    \item \textit{Opacity 0.01 ~\faFilePdf~\faHtml5}.
    \item \textit{Visibility ~\faHtml5}: In HTML, using `\texttt{visibility: hidden;}' (`\texttt{<span style=``visibility: hidden;''>hidden text</span>}') makes the text invisible, but unlike `\texttt{opacity: 0;}`, it also prevents the element from taking up space in the layout. However, the text remains in the HTML source and can still be processed by AI models, crawlers, or screen readers, making it another potential technique for adversarial attacks.
    \item \textit{Vanish ~\faFileWord}: In DOCX, the ``Vanish'' property hides text from being displayed in the document while keeping it in the file structure, allowing it to be processed by text parsers and AI models. This can be applied using Word’s formatting options (`\texttt{<w:vanish/>}' in XML) or by selecting ``Hidden'' in the font settings, making it a potential method for adversarial attacks.
\end{enumerate}

\subsubsection{Zero-Size Font~\faFilePdf~\faFileWord~\faHtml5}
This method reduces the font size of adversarial text to zero or near-zero, making it imperceptible to users but still interpretable by machine learning models. It exploits text rendering differences to introduce hidden prompts or evade detection in AI-driven systems.
We implement the following variants of the attack: 
\begin{enumerate*}
    \item \textit{Size0 ~\faFilePdf~\faHtml5}: font size set to 0, making it invisible. 
    \item \textit{Size0.01 ~\faFilePdf~\faHtml5}: font size set to 0.01, making it nearly invisible. 
    \item \textit{Size1 ~\faFileWord}: font size set to 1, making it very small. This is the minimum size accepted by Word documents. 
\end{enumerate*}

\subsection{Two-fold Effect}
We now present techniques that can be used both for content obfuscation and content injection. 

\subsubsection{Font Poisoning ~\faFilePdf~\faHtml5}
This attack utilizes a custom font to map displayed characters to different underlying characters parsed by document loaders. For example, in a specially designed font, the glyph for the character ``A'' could be associated with the ASCII code for ``B'' (U+0042). This manipulation creates a visual representation that appears as ``A'' to human readers, while the underlying text is actually parsed as ``B'' by automated systems, compromising the integrity of text processing.
The attack effect can be two-fold:
\begin{enumerate}
    \item Semantic disruption, as the attacker might substitute the ASCII representation of glyphs to random characters. For instance, the world ``apple'' could be represented by the ASCII representation ``qwert'', 
    \item Invisible text injection, as the attacker might substitute the ASCII representation of glyphs to target characters. For instance, the world ``apple'' could be represented by the ASCII representation ``drugs'' (note that multiple fonts can be used to map the same glyph to multiple ASCII representations). 
\end{enumerate}
Font poisoning has been introduced in~\cite{markwood2017mirage}, where authors demonstrated the feasibility of PDF content masking against  real-world systems, including Turnitin, Bing, Yahoo!, and DuckDuckGo.

%% file: Sections/99-resultsPhantom.tex
\section{Evaluation}
As part of our contribution, we introduce the \textbf{PhantomText Toolkit}, a modular framework for crafting invisible manipulations in documents (DOCX, HTML, PDF). PhantomText supports the attack strategies introduced in Section~\ref{sec:phantom}. The toolkit is open-source and available at: \url{https://github.com/pajola/PhantomText-Paper} .

This repository contains all material necessary to reproduce our experiments. 
To evaluate the effectiveness of PhantomText, we designed three experiments targeting the end-to-end pipeline of RAG systems.

\begin{itemize}
    \item \textbf{Experiment 1: Poisoning Dataloaders.} We apply all PhantomText techniques to HTML, DOCX, and PDF documents using popular document loader implementations, assessing whether invisible artifacts can survive ingestion and be embedded into the knowledge base. This demostrates that PhantomText techniques can poison the knowledge base by manipulating the content of documents at the data loader stage.
    \item \textbf{Experiment 2: End-to-End RAG Manipulation.} Poisoning the knowledge base does not necessarily guarantee that the RAG system will be impacted in the same way, since the LLM model that generates the output may be resilient to such attacks. For this reason, we study the full end-to-end RAG pipeline, to determine if the invisible manipulations in the poisoned documents influence the retriever and the generative model. 
    \item \textbf{Experiment 3: CIA triad-oriented RAG attacks.} 
    Finally, we demonstrate how the techniques we described can be used by malicious actors to perform the attacks described in Section~\ref{sec:threat-landscape}. We composed targeted attacks using the most effective techniques from the toolkit, showing that PhantomText enables subtle, high-impact attacks.
\end{itemize}
In the following sections, we first present the research question that motivates our experiments, then detail the methodology we followed to perform the experiments, and finally report the results we obtained, with an answer for the research question.

\subsection{Exp1: Poisoning Data Loaders}\label{sec:Experiment_1}
\subsubsection{Overview}
Here, we assess the efficacy of PhantomText in poisoning the RAG systems’ knowledge base by targeting the document parsing frameworks’ Data Loader components. It tests if invisible manipulations, such as obfuscation and injection, can compromise ingestion of HTML, DOCX, and PDF documents; results show PhantomText effectively embeds malicious artifacts to poison RAG knowledge bases.

\begin{bluetextbox}
{\scriptsize \faQuestionCircle} \textbf{\textit{Research Question:}} 
Are popular Document Loaders vulnerable to PhantomText toolkit content obfuscation and injection techniques? 
\end{bluetextbox}

\subsubsection{Experimental Settings}\label{Experimental_Setting_01}
\paragraph{Document Loaders}
We tested the following frameworks: Docling~\cite{docling, docling2}, Haystack, LangChain, LlamaIndex, and LLMSherpa.\footnote{We provide the list of software in the Appendix~\ref{app:los}} These frameworks were selected based on their popularity, trending status on GitHub, and widespread adoption in RAG development. For each framework, we focused on the document parsers responsible for handling different file formats, i.e. PDF, HTML, and DOCX. In total, 21 different parsers across the selected frameworks were evaluated. Specifically, for Docling we used the Default parser, the PyPDFium parser, the HTML parser, the DOCX parser, and the OCR parser; for Haystack, the PyPDF parser, the PDFMiner parser, the HTML parser, and the DOCX parser were evaluated; for LlamaIndex, we assessed the LlamaParse, SimpleDirectoryReader, and HTMLTagReader; for LangChain, the PyPDF (with and without OCR), PDFPlumber, PyPDFium2, PyMUPDF (with and without OCR), UnstructuredHTML, BSHtml, and Docx2txt parsers were employed; finally, the LLMSherpa Default parser was evaluated.\footnote{The OCR we used was tesseract for all the loaders.}

\paragraph{Dataset}
We created a custom dataset to test PhantomText's ability to obfuscate and inject content into DOCX, HTML, and PDF files. The dataset is created by utilizing 100 random samples from Amazon Reviews'23~\cite{hou2024bridging}. Then, starting from each sample, we generate a poisoned document as follows: for \textit{content obfuscation} techniques, a random word is selected within the document text and replaced with the injection counterpart; in the \textit{content injection}, we insert a random word not present in the document.\footnote{The random word is picked from this list: \url{https://www.kaggle.com/datasets/rtatman/english-word-frequency}.} This procedure is repeated for all the various techniques we presented in Section~\ref{sec:phantom}, generating a dataset of 4200 documents. 
In other words, our dataset now contains those 100 original samples repeated with different injection techniques (e.g., 100 documents are obfuscated with \textit{zero-width character mask1}, 100 with \textit{zero-width character mask2}, and so on). 
We test each document loader configuration with such a dataset, resulting in a total of 35,900 evaluations.
\par
It is important to clarify that our evaluation focuses on the robustness of document ingestion pipelines—software components responsible for parsing and encoding input documents prior to language model inference—rather than on the performance or generalization of ML models themselves. These preprocessing algorithms are domain-agnostic, operating at a syntactic and structural level regardless of semantic content or text domain. Thus, Amazon Reviews provide a controlled yet realistic dataset, suitable for systematically evaluating injection attacks on software vulnerabilities.

\paragraph{Evaluation}
A single evaluation is considered successful if, in the \textit{Content Obfuscation} scenario, the target word was completely absent from the parsed text, and in the \textit{Content Injection} scenario, the synthetic word was correctly present in the parsed output. The \textbf{Attack Success Rate (ASR)} is defined as the ratio between the number of successful tests and the total number of attempts, i.e.,
\[
\text{ASR} = \frac{\text{Succes}}{\text{Succes} + \text{Failed}}.
\]

\subsubsection{Results}
Due to space limitations, a detailed table presenting the attack success rate for each configuration is provided exclusively in the public repository. Below, we present an overview of these findings.

\paragraph{Overall Trends}
PhantomText Toolkit demonstrated an overall success rate of 74.4\%, indicating a significant degree of effectiveness in compromising the integrity of RAG's document loaders. Furthermore, out of 375 tests conducted, 238 yield a success rate over 95\%.  
When analyzed by attack family, the performance shows notable distinctions between the two primary strategies. The content obfuscation and injection families achieved, respectively, a success rate of 76.7\% and 72.4\%. 

\begin{figure}[!htpb]
     \centering
     \begin{subfigure}[b]{0.8\linewidth}
         \centering
         \includegraphics[width=\textwidth]{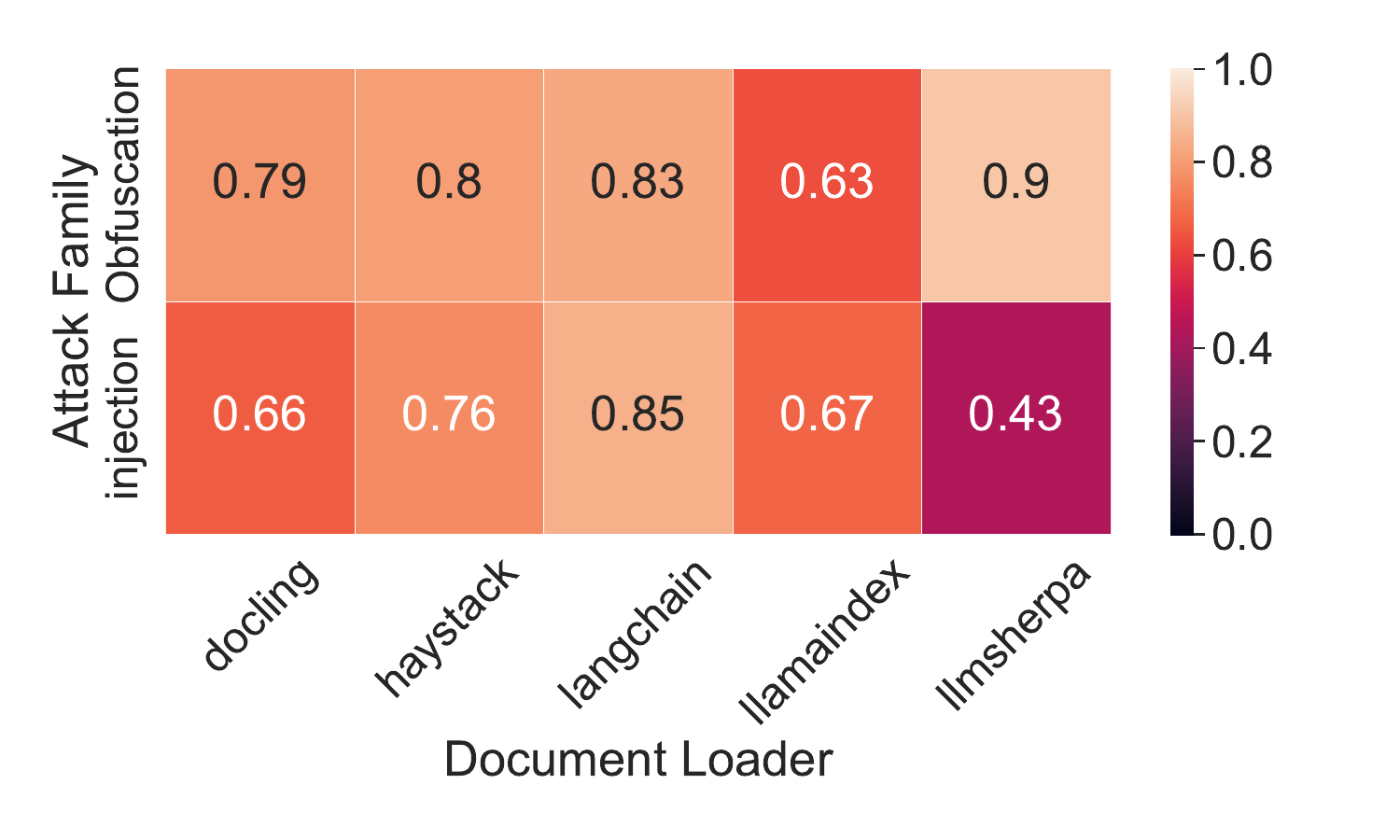}
         \caption{Document Loaders}
         \label{fig:docLoaderASR}
     \end{subfigure}
     \hfill
     \begin{subfigure}[b]{0.8\linewidth}
         \centering
         \includegraphics[width=\textwidth]{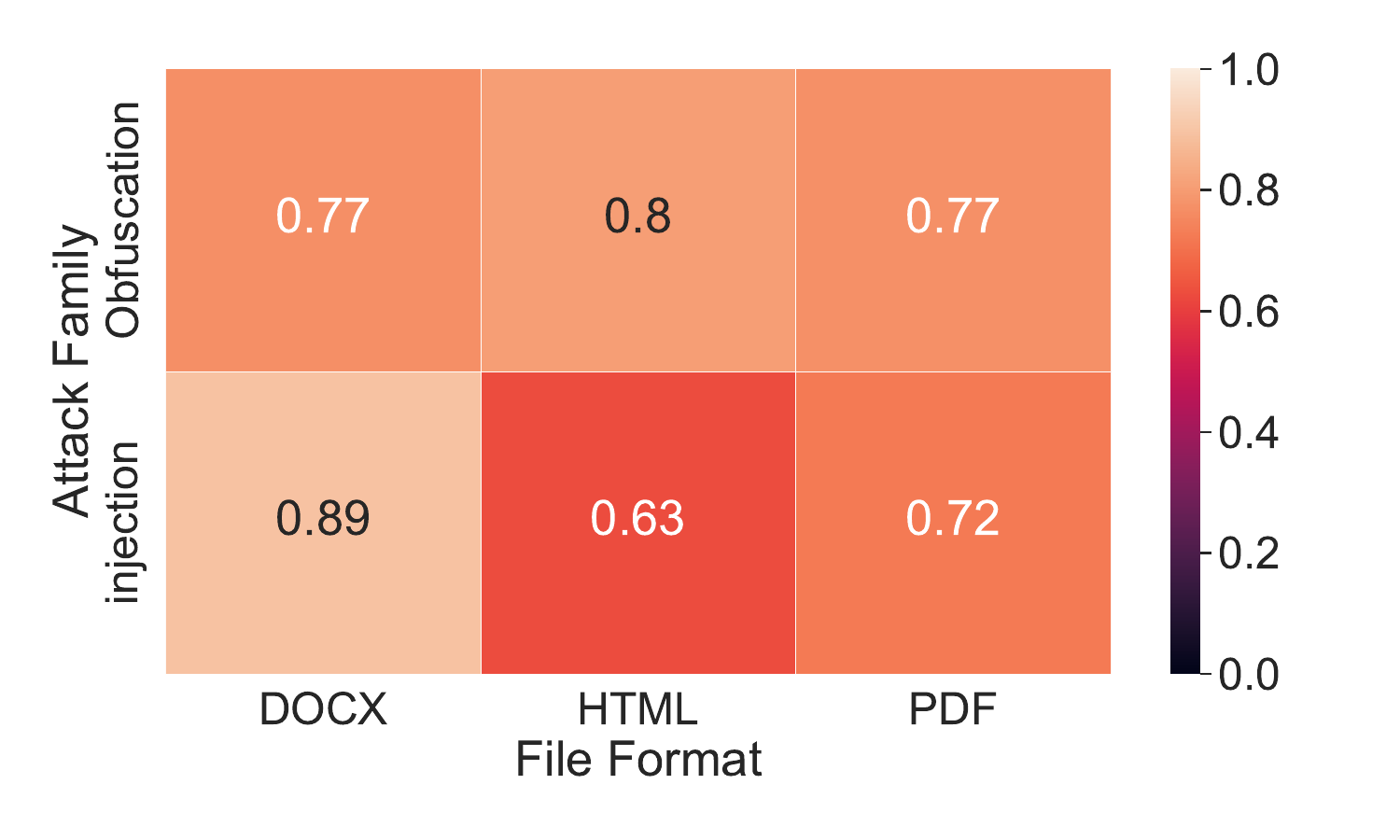}
         \caption{File Format}
         \label{fig:fileFormatASR}
     \end{subfigure}
     \hfill
        \caption{Attack Success Rate at varying poisoning families --  from both \textit{Semantic Corruption via Obfuscation} (in blue) and \textit{Invisible Content Injection} (in orange) families  -- across different data loaders (left) and file format (right).}
        \label{fig:ASR_local-trends}
\end{figure}

\paragraph{Document Loaders Level}
Figure~\ref{fig:docLoaderASR} visualizes the varying effectiveness of the attacks across different data loaders. With an ASR above 0.6 for all loaders, the attack proves effective in manipulating document processing pipelines. Langchain shows the highest ASR, indicating significant vulnerability, while LlamaIndex exhibits the lowest ASR, demonstrating more resistance. Docling, LLMSherpa, and Haystack fall in the mid-range, showing moderate susceptibility. Overall, the results highlight that the attack is effective across all tested loaders, underscoring the need for improved defenses against such adversarial manipulations.

\paragraph{File Format}
Figure~\ref{fig:fileFormatASR} illustrates the effectiveness of the Text Injection attack across different file formats. All formats exhibit ASRs above 0.6, indicating that the attack is effective in each case. DOCX shows the highest ASR at 0.89, suggesting a high level of vulnerability. Overall, the results highlight that while the attack is effective across all file formats, DOCX is the most vulnerable, and PDF offers the best defense among the tested formats.

\begin{figure}[!htpb]
     \centering
     \begin{subfigure}[b]{0.99\linewidth}
         \centering
         \includegraphics[width=\textwidth]{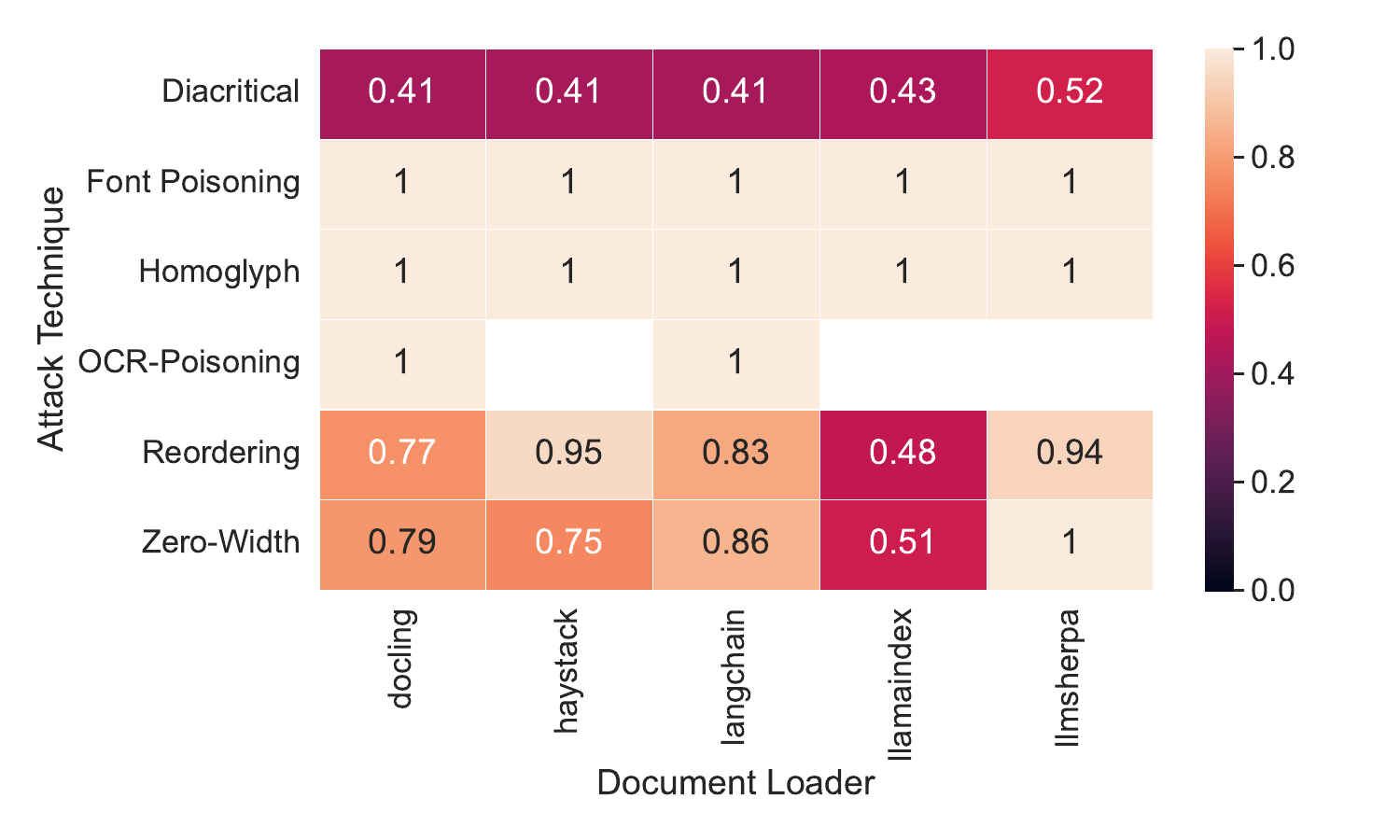}
         \caption{Content Obfuscation}
     \end{subfigure}
     \hfill
     \begin{subfigure}[b]{0.99\linewidth}
         \centering
         \includegraphics[width=\textwidth]{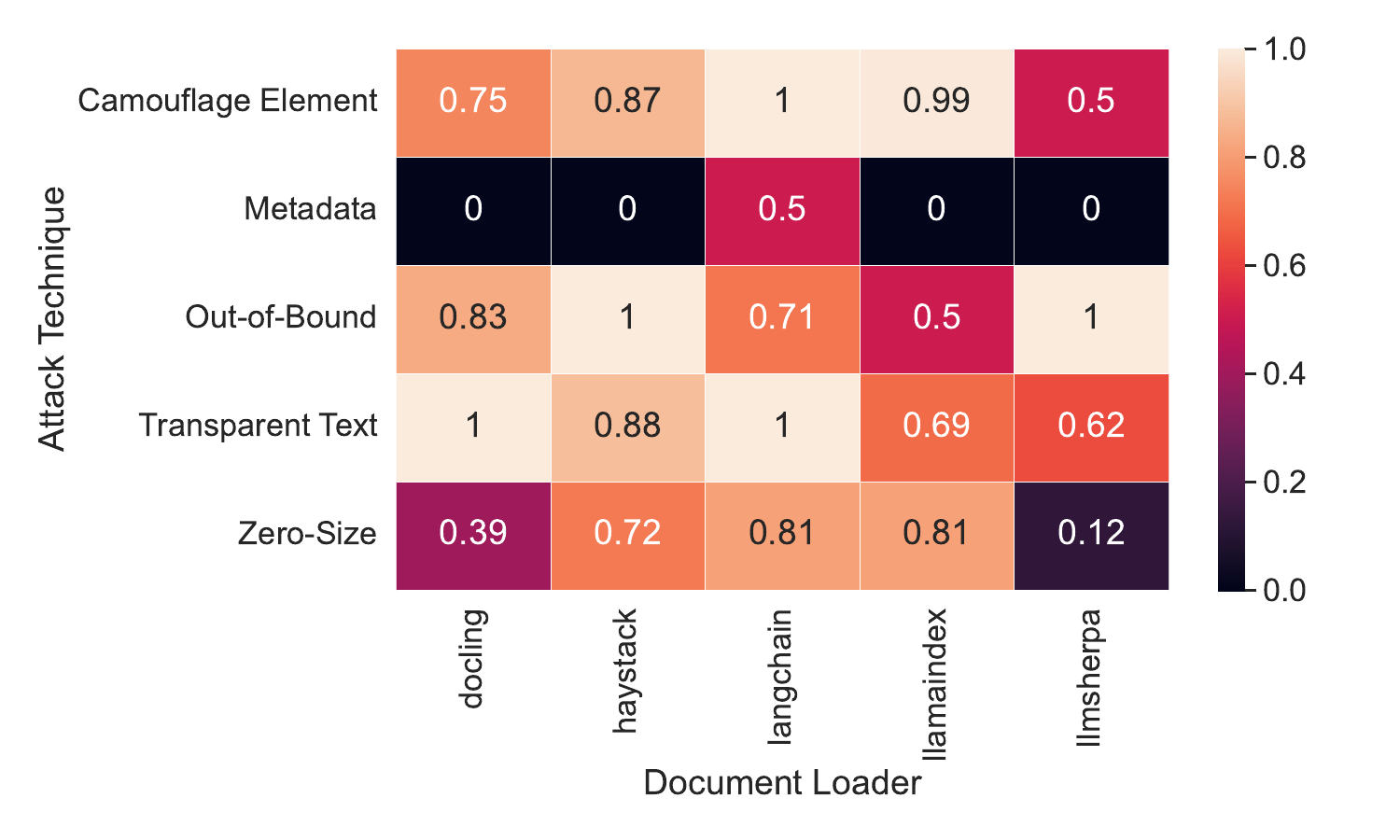}
         \caption{Content Injection}
     \end{subfigure}
     \hfill
        \caption{Attack Success Rate at varying poisoning families (Content Obfuscation and Injection).}
        \label{fig:ASR_local_technique}
\end{figure}

\paragraph{Success Rate of each Attack Technique}
We conclude by analyzing the impact of each poisoning technique on document loaders, as shown in Figure~\ref{fig:ASR_local_technique}. The results reveal that not all techniques are equally effective. For example, \textit{font poisoning} and \textit{homoglyph characters} consistently disrupt all data loaders tested, achieving a 100\% attack success rate ($ASR = 1.0$). In contrast, \textit{diacritical injection} shows limited effectiveness, with an average $ASR$ of approximately 0.4. \textit{Metadata injection} proves to be largely ineffective across the board, except for Langchain, where it reaches a moderate $ASR$ of 0.5.
The analysis also highlights significant variability in how different data loaders respond to specific attacks. For example, \textit{ out-of-bound text} is particularly effective against Docling ($ASR = 0.83$) but less so against LlamaIndex ($ASR = 0.5$). In contrast, LlamaIndex is more vulnerable to \textit{zero-size font} ($ASR = 0.81$), while Docling demonstrates greater resistance to the same technique ($ASR = 0.39$).
These findings indicate that attack effectiveness is highly technique-dependent and that document loaders exhibit varying levels of resilience. Therefore, the choice of technique must be tailored to the specific target in order to maximize adversarial impact.

\begin{redtextbox}
{\scriptsize \faChartLine } \textbf{\textit{Yes}, Popular Document Loaders are severely vulnerable.} 
We observed that both injection and obfuscation techniques persist through preprocessing and are embedded into the vector store, enabling successful poisoning across multiple formats.
\end{redtextbox}

\subsection{Exp2: End-to-End RAG Manipulation}\label{sec:Experiment_2}
\subsubsection{Overview}
Building on the results from Experiment 1, we explore whether a poisoned knowledge base affects the behavior of RAG systems. Specifically, we focus on determining whether the invisible manipulations introduced into the documents during ingestion can propagate through the retriever and generative model components of popular RAG frameworks. While the knowledge base is successfully poisoned in Experiment 1, the impact on system behavior is more nuanced, with some systems showing greater resilience than others. For simplicity, we focus only on PDF documents; however, the results can be generalized to other document formats, such as DOCX and HTML. This experiment demonstrates that document poisoning does not always guarantee that the entire RAG pipeline will be affected, and shows which techniques are the most effective in attacking different RAG pipelines.

\begin{bluetextbox}
{\scriptsize \faQuestionCircle} \textbf{\textit{Research Question:}} 
Can PhantomText toolkit content obfuscation and injection techniques affect end-to-end RAG pipelines? Which techniques are the most effective?
\end{bluetextbox}

\subsubsection{Experimental Settings}

\paragraph{RAG Configuration}
In this work we tested different end-to-end RAG frameworks, with a total of 6 different setups ($M = 6$):
\begin{itemize}
    \item \textbf{White-box} setting. Open-source frameworks that give us the capability to fully access the internal components, including the retriever and the language model. We utilized three distinct LLMs: Llama 3.2 (3B), Gemma 3 (27B), and Deep Seek R1. The RAG utilizes a local retriever. We locally implement 3 distinct RAG utilizing the official LangChain documentation and a Chroma retriever\footnote{\url{https://python.langchain.com/docs/tutorials/rag/}}. A detailed description of our setup is reported in Appendix~\ref{app:rag-det}.
    \item \textbf{Black-box} setting. It involves using pre-built RAG systems via APIs or web interfaces. We tested two black-box RAGs: OpenAI assistants (gpt-4o and o3-mini) and NotebookLM, based on Google's Gemini 2.0 Flash.
\end{itemize}
Two LLMs -- Deep Seek R1 and OpenAI o3-mini -- were trained for \textit{reasoning} responses. 
\par
In order to maximize fairness and ensure the validity of our RAG pipeline, for 5 out of 6 RAGs, we integrate a custom \textit{system prompt} sourced directly from the OpenAI Playground interface, as reported in Appendix~\ref{app:rag-det}. This prompt is notably more rigorous than Langchain’s default version; we adopted it to better simulate real-world scenarios and minimize bias in our experiments. The only exception is NotebookLM, which operates as a fully black-box system and does not allow users to customize the system prompt.

\paragraph{Dataset}
The aim of this experiment is to understand if it is possible to affect end-to-end RAGs with PhantomText techniques, and to do so we deploy a straightforward test. We start from a benign document containing information about a fictional company,  ensuring that RAG responses are not conditioned by the prior knowledge of the LLMs. 
Then, we produce 14 poisoned documents ($D=14$) - one for each PDF PhantomText technique - in two fashions: for content obfuscation families, we use the techniques to hide specific information, while for the content injection, we add a new concept related a fictional competitor company. 

\paragraph{Evaluation}
We first evaluate the base quality of the RAG systems to ensure the validity of the experiment. We utilize a benign document and we generate queries to ask about the fictional company and its competitor ($Q = 4$). We test the 6 RAG systems against these queries, and verify that they are able to correctly answer. We repeat each query 10 times, to reduce randomness ($R = 10$). In total, we conduct $Q \times R \times M = 240$ tests.
Then, we analyze the effectiveness of PhantomText toolkit, and its ability to obfuscate and inject information, by repeating the queries with poisoned documents. In this case, our total tests are $D \times R \times M = 840$ tests.
All responses are manually evaluated to check whether the attack was successful.

\subsubsection{Results}
\begin{figure}
    \centering
    \includegraphics[width=0.99\linewidth]{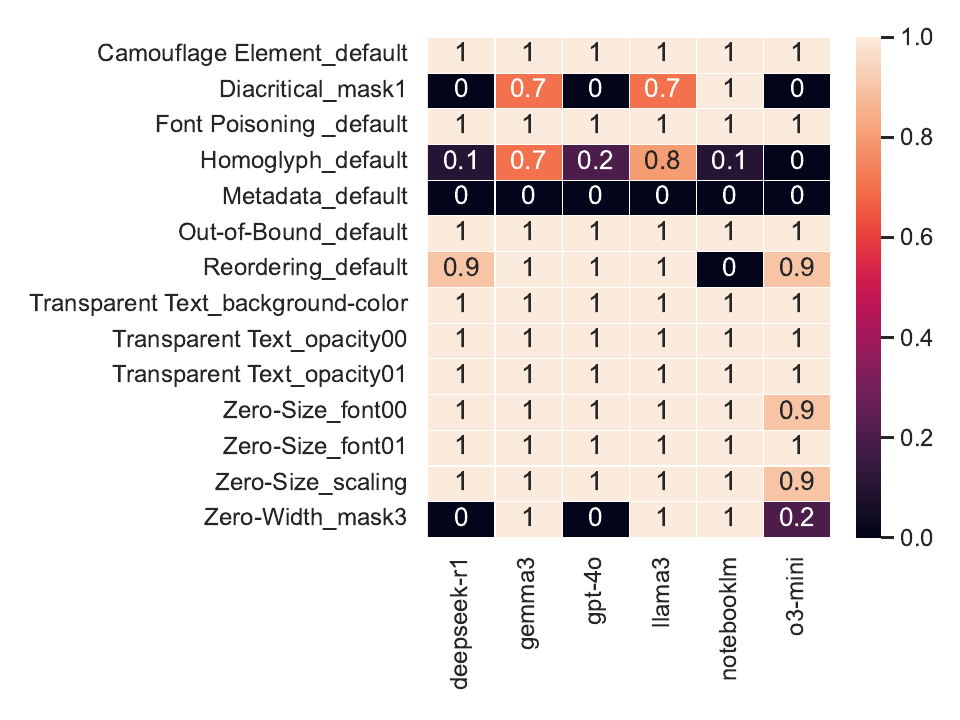}
    \caption{\textit{Experiment 2 -- End-to-End RAG Manipulation}.}
    \label{fig:exp2-heat}
\end{figure}

The quality assurance test resulted in 100\% of accuracy, meaning that the RAG systems were able to correctly answer the queries based on the contents of the provided documents. 
As for the attack techniques, we report our results in Figure~\ref{fig:exp2-heat}. We note that distinct patterns emerge across the various techniques. Most attacks implemented in PhantomText -- such as \textit{camouflage element} and \textit{font poisoning} -- consistently lead to effective RAG manipulation ($ASR = 1.0$) across all tested RAG systems, in both white-box and black-box settings. However, a small number of techniques such as \textit{metadata} show no effect ($ASR = 0$). Finally, \textit{reordering} and \textit{homoglyphs} generate a range of outcomes, with some models achieving close to full success, while others, exhibiting reduced or null susceptibility. These findings highlight the varying degrees of vulnerability of different AI models to specific attacks. Due to space limitations, a detailed table breaking down the attack evaluation is provided only in the public repository.


\begin{redtextbox}
{\scriptsize \faChartLine } \textbf{\textit{Yes}, RAG systems can be affected by PhantomText techniques.} 
Not all PhantomText techniques are equally effective for RAG end-to-end systems; most are always effective (e.g., font poisoning), others consistently fail (e.g., metadata), and some are effective only on specific RAG systems (e.g., homoglyphs).
\end{redtextbox}

\subsection{Exp3: CIA triad-oriented RAG attacks}\label{sec:Experiment_3}
\subsubsection{Overview}
We finally apply the most effective PhantomText techniques to craft targeted attacks that exploit vulnerabilities in RAG systems presented in Section~\ref{sec:threat-landscape}, with a focus on the \emph{Confidentiality}, \emph{Integrity}, and \emph{Availability} (CIA) triad. Using popular RAG frameworks, we demonstrate how invisible manipulations can lead to attacks such as leaking sensitive information, altering retrieved facts, and degrading system functionality. While we focus on PDF documents for simplicity, these techniques can be transferred to other document formats like DOCX and HTML. These attacks show that the combination of invisible content and subtle manipulations can be leveraged to bypass typical detection mechanisms, posing significant risks to both the integrity and availability of the system.

\begin{bluetextbox}
{\scriptsize \faQuestionCircle} \textbf{\textit{Research Question:}} 
Can PhantomText toolkit techniques be leveraged to perform the attacks described in Section~\ref{sec:threat-landscape}, affecting the confidentiality, integrity, and availability of state-of-the-art RAG systems?
\end{bluetextbox}

\subsubsection{Experimental Settings}
\paragraph{Datasets}
We conduct attacks on the same RAG systems presented in Section~\ref{sec:Experiment_2}.
For each of the nine attack scenarios listed in Section~\ref{sec:threat-landscape}, we crafted a series of datasets with different scenarios. For instance, when performing \textit{sensitive data disclosure}, we designed a dataset containing sensitive information. As a representative case from the public repository (Appendix~\ref{app:exp3-details}), we detail one example to illustrate the attack's mechanics and impact.
In general, we defined 4 distinct dataset per scenario, with a varying number of documents ($N = \{1, 100\}$) and type of content (i.e., poisoned and unpoisoned). Unpoisoned datasets are utilized as a control group to verify the correct execution of RAGs (i.e., if the RAG is able to answer the query without any tampering), while poisoned datasets are utilized to assess PhantomText attacks. In particular, for poisoned dataset, we are interested in two cases: 
\begin{itemize}
    \item \textbf{Single Poisoned Document}. The knowledge base contains only a single poisoned document. This scenario is designed to assess the maximum effectiveness of a specific attack when the adversary has complete control over the information source available to the RAG system.
    \item \textbf{100 Poisoned Documents}. The knowledge base contains all 100 documents (i.e. 99 legitimate and 1 poisoned). This scenario aimed to evaluate the attack performance in a more realistic setting where the attacker's ability to manipulate the knowledge base is limited.
\end{itemize}
For each dataset and attack scenario, we designed ad-hoc queries. 

\input{Tables/exp3-tab}

\paragraph{Evaluation}
We evaluate each attack scenario across the 6 RAG systems, using an LLM-as-a-judge to label the answers given by the RAG systems as either successful attacks or failures. The only exceptions are attacks a3, which was manually analyzed, and a9, where we used a simple substring match to look for leaked information in the responses. More details about our judging approach for the rest of attacks are reported in Appendix~\ref{app:judge}.
Each experiment is repeated 10 times to reduce randomness due to the non-deterministic nature of LLMs (with the exception of \textit{pipeline failure} where we execute the test one time since we need to understand if the end-to-end RAG ``crashes'' and does not involve LLMs).
For each attack, we report the number of successful executions out of 10, except for \textit{pipeline failure} (single test) and \textit{reasoning overload}, where we report the average increase factor in tokens used during reasoning.

\subsubsection{Results} 
The results of the attacks are presented in Table~\ref{tab:Exp3}, and highlight distinct vulnerabilities of RAG systems when targeted by PhantomText attacks in both white-box and black-box settings. A consistent pattern of high attack success rate is observed in several failure modes, especially in white-box scenarios. This indicates a critical lack of robustness in these systems against semantic disruption techniques that can affect confidentiality, integrity, and availability. Interestingly, even advanced models like deepseek-r1 are very susceptible to all executed attacks. 
\par
Black-box models display a more heterogeneous vulnerability profile. For instance, \textbf{NotebookLM} demonstrated strong resilience to several attack vectors, with a 0\% success rate in attacks like \textit{factual distortion}, \textit{empty statement response}. On the other hand, commercial models such as \textbf{gpt-4o} and \textbf{o3-mini} were significantly more vulnerable, with high success rates (up to 10/10) in many categories. Specifically, attacks targeting \textbf{o3-mini} were always successful for 9 out of 16 scenarios. The results suggest that while some black-box models may benefit from stronger internal filtering or post-processing mechanisms for some attacks, they remain in general as vulnerable as their white-box counterparts. Importantly, the \textit{reasoning-overload} attack, which measures token inflation, showed a very noticeable increase in response length for certain models (e.g., 4.9$\times$ in \textbf{o3-mini}), indicating that resource abuse or inefficiency can also be induced adversarially.

A separate discussion can be made about the \textit{pipeline failure} attack, since the results show that only our white-box RAG pipelines where vulnerable to this attack. First of all, we need to point out that we only tested a very simple attack vector, i.e. a very big PDF document, which is very easy to detect. More sophisticated attacks may bypass the detection of even the black-box RAG systems. Additionally, the results may be interpreted as if the problem lies in our implementation of the RAG pipeline. However, we want to highlight the fact that we followed the implementation suggested in the official Langchain documentation. It is likely that unexperienced developers follow that same documentation, resulting in a faulty RAG pipeline that crashes with such a simple attack.  

\begin{redtextbox}
{\scriptsize \faChartLine } \textbf{\textit{Yes}, Invisible injection can affect confidentiality, integrity, and availability of state-of-the-art RAG systems.} 
Our demonstration highlights the need to carefully design these systems with cyber-secure principles.
\end{redtextbox}

%% file: Tables/exp3-tab.tex
\begin{table*}[!htpb]
\centering
\resizebox{0.8\textwidth}{!}{
\begin{tabular}{l|cc|cc|cc|cc|cc|cc|}
& \multicolumn{6}{c|}{\textit{White-box}} & \multicolumn{6}{c|}{\textit{Black-box}} \\
 & \multicolumn{2}{c}{\textbf{deepseek-r1}} & \multicolumn{2}{c}{\textbf{gemma3}} & \multicolumn{2}{c|}{\textbf{llama}} & \multicolumn{2}{c}{\textbf{NotebookLM}} & \multicolumn{2}{c}{\textbf{gpt-4o}} & \multicolumn{2}{c|}{\textbf{o3-mini}} \\ & 1 & 100 & 1 & 100 & 1 & 100 & 1 & 100 & 1 & 100 & 1 & 100 \\
\midrule
a1-pipeline-failure & \cellcolor[RGB]{255,160,160} 1/1 & \cellcolor[RGB]{255,160,160} 1/1 & \cellcolor[RGB]{255,160,160} 1/1 & \cellcolor[RGB]{255,160,160} 1/1 & \cellcolor[RGB]{255,160,160} 1/1 & \cellcolor[RGB]{255,160,160} 1/1 & \cellcolor[RGB]{144,238,144} 0/1 & \cellcolor[RGB]{144,238,144} 0/1 & \cellcolor[RGB]{144,238,144} 0/1 & \cellcolor[RGB]{144,238,144} 0/1 & \cellcolor[RGB]{144,238,144} 0/1 & \cellcolor[RGB]{144,238,144} 0/1 \\
a2-reasoning-overload & \cellcolor[RGB]{255,160,160} 1.25 & \cellcolor[RGB]{255,160,160} 2.04 & / & / & / & / & / & / & / & / & \cellcolor[RGB]{255,160,160} 4.90 & \cellcolor[RGB]{255,160,160} 3.48 \\
a3-unreadable-output & \cellcolor[RGB]{222,227,163} 5/10 & \cellcolor[RGB]{222,227,163} 4/10 & \cellcolor[RGB]{144,238,144} 0/10 & \cellcolor[RGB]{144,238,144} 0/10 & \cellcolor[RGB]{144,238,144} 0/10 & \cellcolor[RGB]{163,235,149} 0/10 & \cellcolor[RGB]{144,238,144} 0/10 & \cellcolor[RGB]{249,186,163} 8/10 & \cellcolor[RGB]{163,235,149} 9/10 & \cellcolor[RGB]{163,235,149} 1/10 & \cellcolor[RGB]{183,232,154} 9/10 & \cellcolor[RGB]{144,238,144} 5/10 \\
a4-empty-statement-response & \cellcolor[RGB]{255,160,160} 10/10 & \cellcolor[RGB]{255,160,160} 10/10 & \cellcolor[RGB]{255,160,160} 10/10 & \cellcolor[RGB]{255,160,160} 10/10 & \cellcolor[RGB]{255,160,160} 10/10 & \cellcolor[RGB]{255,160,160} 10/10 & \cellcolor[RGB]{144,238,144} 0/10 & \cellcolor[RGB]{144,238,144} 0/10 & \cellcolor[RGB]{249,186,163} 8/10 & \cellcolor[RGB]{249,186,163} 8/10 & \cellcolor[RGB]{255,160,160} 10/10 & \cellcolor[RGB]{249,186,163} 8/10 \\
a5-ambiguous-output & \cellcolor[RGB]{255,160,160} 10/10 & \cellcolor[RGB]{255,160,160} 10/10 & \cellcolor[RGB]{255,160,160} 10/10 & \cellcolor[RGB]{255,160,160} 10/10 & \cellcolor[RGB]{255,160,160} 10/10 & \cellcolor[RGB]{255,160,160} 10/10 & \cellcolor[RGB]{255,160,160} 10/10 & \cellcolor[RGB]{255,160,160} 10/10 & \cellcolor[RGB]{255,160,160} 10/10 & \cellcolor[RGB]{255,160,160} 10/10 & \cellcolor[RGB]{255,160,160} 10/10 & \cellcolor[RGB]{255,160,160} 10/10 \\
a6-bias-injection & \cellcolor[RGB]{255,160,160} 10/10 & \cellcolor[RGB]{255,160,160} 10/10 & \cellcolor[RGB]{255,160,160} 10/10 & \cellcolor[RGB]{255,160,160} 10/10 & \cellcolor[RGB]{252,173,162} 9/10 & \cellcolor[RGB]{255,160,160} 10/10 & \cellcolor[RGB]{144,238,144} 0/10 & \cellcolor[RGB]{252,173,162} 9/10 & \cellcolor[RGB]{247,198,165} 7/10 & \cellcolor[RGB]{252,173,162} 9/10 & \cellcolor[RGB]{255,160,160} 10/10 & \cellcolor[RGB]{255,160,160} 10/10 \\
a7-factual-distortion & \cellcolor[RGB]{255,160,160} 10/10 & \cellcolor[RGB]{255,160,160} 10/10 & \cellcolor[RGB]{255,160,160} 10/10 & \cellcolor[RGB]{255,160,160} 10/10 & \cellcolor[RGB]{255,160,160} 10/10 & \cellcolor[RGB]{255,160,160} 10/10 & \cellcolor[RGB]{144,238,144} 0/10 & \cellcolor[RGB]{144,238,144} 0/10 & \cellcolor[RGB]{255,160,160} 10/10 & \cellcolor[RGB]{255,160,160} 10/10 & \cellcolor[RGB]{255,160,160} 10/10 & \cellcolor[RGB]{255,160,160} 10/10 \\
a8-outdated-knowledge & \cellcolor[RGB]{255,160,160} 10/10 & \cellcolor[RGB]{255,160,160} 10/10 & \cellcolor[RGB]{255,160,160} 10/10 & \cellcolor[RGB]{255,160,160} 10/10 & \cellcolor[RGB]{255,160,160} 10/10 & \cellcolor[RGB]{255,160,160} 10/10 & \cellcolor[RGB]{255,160,160} 10/10 & \cellcolor[RGB]{255,160,160} 10/10 & \cellcolor[RGB]{252,173,162} 9/10 & \cellcolor[RGB]{249,186,163} 8/10 & \cellcolor[RGB]{255,160,160} 10/10 & \cellcolor[RGB]{255,160,160} 10/10 \\
a9-sensitive-data-disclosure & \cellcolor[RGB]{255,160,160} 10/10 & \cellcolor[RGB]{255,160,160} 10/10 & \cellcolor[RGB]{255,160,160} 10/10 & \cellcolor[RGB]{255,160,160} 10/10 & \cellcolor[RGB]{144,238,144} 0/10 & \cellcolor[RGB]{144,238,144} 0/10 & / & / & \cellcolor[RGB]{144,238,144} 0/10 & \cellcolor[RGB]{144,238,144} 0/10 & \cellcolor[RGB]{249,186,163} 8/10 & \cellcolor[RGB]{252,173,162} 9/10 \\
\bottomrule
\end{tabular}%
}
\caption{{\textit{Experiment 3 -- Real-World Exploitation Across RAG Attack Scenarios.} PhantomText attack success in executing different attacks to different RAG end-to-end systems, with 1 and 100 documents. Note that we report the attack success rate for each row, while for the attack ``reasoning overload'' we report the average factor of increase in the number of tokens. }}
\label{tab:Exp3}
\end{table*}

%% file: Sections/10-Discussions.tex
\section{Defenses}
Although the presented injection strategies are diverse, many share syntactic or formatting patterns that can be detected using simple heuristics. Several attacks introduce structural or visual anomalies—imperceptible to humans but detectable via shallow pattern analysis or normalization. For example, diacritical mark stacking produces abnormal Unicode sequences that deviate from language norms. Statistical checks on combining marks or character frequencies can flag these. Homoglyph substitutions—though visually deceptive—leave traces in the form of anomalous Unicode blocks (e.g., Cyrillic in Latin text), which can be neutralized through character set filtering or canonical Unicode normalization (e.g., NFC).
\par
Zero-width characters are rare in natural text and can be detected by scanning for a fixed set of Unicode code points. A blanket removal of such characters is often safe and effective. Bidirectional control characters (used in reordering attacks) introduce RTL/LTR switches that can be flagged by inspecting Unicode identifiers, especially when occurring mid-word or mid-token. Invisible injections using formatting tricks—e.g., transparent text, vanished styles, or zero-size fonts—can be detected via document parsing, targeting attributes like color, opacity, or tags (e.g., \texttt{<w:vanish/>}). Similarly, out-of-bound injections can be flagged through bounding-box analysis of rendered layout or structural metadata.
\par
While these defenses are not foolproof, lightweight sanitization can mitigate many attacks at low cost. For more robust protection, advanced techniques like OCR-based pipelines~\cite{boucher2022bad} offer resilience against invisible perturbations but incur computational overhead and potential transcription errors. We also explored this OCR-based ingestion module (we tried two popular open-source OCR, i.e. EasyOCR and Tesseract), which extracts text from rendered document images, effectively bypassing many parser-level exploits. This approach blocks more 90\% invisibility-based attacks, such as transparent or out-of-bound text, substantially reducing the attack success rate. However, OCR incurs significant computational overhead and may introduce transcription errors, limiting its practicality in high-throughput environments. Importantly, OCR is not a silver bullet: certain attacks, including diacritical manipulations and zero-size font injections, can still evade OCR-based defenses. Therefore, OCR should be considered a strong first layer in a multi-layered defense strategy rather than a standalone solution.
These defensive insights and empirical results will be integrated into the revised manuscript to provide a balanced and actionable security perspective.

\section{Related Works}

\paragraph{Imperceptible Characters against NLP}
Imperceptible perturbations to text have emerged as a powerful attack vector against NLP systems. Pajola et al.~\cite{pajola2021fall} introduced Zero-Width (ZeW) attacks, injecting invisible Unicode characters to bypass classifiers by major platforms like Google and Amazon. This was expanded~\cite{boucher2022bad} to include wide class of black-box attacks using encoding-level manipulations—zero-width, homoglyphs, and reordering— to break commercial NLP systems.
Such attacks are also effective in subverting search engines and LLMs by stealthily altering queries and content, impacting retrieval, summarization, and ranking.~\cite{boucher2023boosting}
Our work took inspiration by these techniques, applying invisible injections to poison the document corpus of RAG systems.

\paragraph{Security of LLMs and RAG Systems}
LLMs have shown impressive capabilities but also raise significant security concerns. Yao et al.~\cite{yao2024survey} categorize the dual nature of LLMs as “The Good” (e.g., code analysis, privacy protection), “The Bad” (e.g., offensive misuse), and “The Ugly” (e.g., inherent vulnerabilities). LLMs have been shown to be vulnerable to black-box jailbreak prompts that can bypass state-of-the-art safety measures~\cite{mehrotra2024tree}.
RAG systems have also been studied, with attacks that can be used to create retrieval backdoors in RAG sytems(TrojRAG~\cite{badrag}). Other attacks include prompt injections through documents~\cite{chaudhari2024phantom} and knowledge corruption with specific question/answer pairs~\cite{poisonedrag}.
These works highlight the urgent need for robust defenses across both LLMs and their augmented systems.

\section{Conclusion}

Our analysis reveals significant vulnerabilities in RAG systems’ data loading phase, stemming from widespread neglect of input sanitization. These weaknesses allow stealthy injection attacks that compromise the integrity of downstream language model outputs. Addressing these issues requires the disciplined application of established security practices to document ingestion pipelines.
Furthermore, as AI systems increasingly rely on retrieved content for complex, multi-step reasoning, similar vulnerabilities may propagate and amplify across workflows. Future defenses must therefore advance beyond static sanitization to include context-aware filtering, provenance verification, and runtime anomaly detection to secure next-generation AI applications.

%% file: Sections/Appendix.tex
\clearpage
\section{Appendix}
The full appendix is in the public repository due to space limits.
\subsection*{Ethical Statement}

All experiments were conducted in accordance with ethical research guidelines. No proprietary or production systems were harmed or targeted during this study. The vulnerabilities identified in third-party open-source and commercial tools and libraries were responsibly disclosed to the respective parties prior to publication, following coordinated disclosure practices. Our goal is to raise awareness of these attack surfaces and promote the development of more secure and resilient RAG pipelines.

\subsection{LLM-as-a-judge}\label{app:judge}
For the experiment in Section~\ref{sec:Experiment_3}, due to the large number of tests, we used an LLM-as-a-Judge approach, where one LLM evaluates another’s output. This enables automated annotation of vast data and is increasingly popular~\cite{zheng2023judging}.

We used \texttt{GPT-4o-mini} as the judge model to evaluate answers from various RAG systems, determining if they contained specific characteristics indicative of successful attacks. The full definition of the prompts for all attacks can be found at \url{https://drive.google.com/drive/folders/1Ee4HF1MQOO1jQt_8uuwDrj79qrPb2iX8?usp=drive_link} (due to responsible disclosure, right now it is a google drive folder, and it will be replaced with a public repositoryu upon paper publication).
To evaluate this approach, we randomly sampled 100 RAG outputs for manual analysis and compared them to the LLM judge’s evaluation. Only 3 samples were mislabeled. The 99\% confidence Wilson Score Interval for accuracy is $[0.8891, 0.9923]$.

To evaluate the performance of this approach, we randomly sampled 100 RAG outputs, which we manually analyzed. We compared our evaluation to the evaluation given by the LLM judge, and out of 100 samples only 3 samples were mislabeled. We computed the Wilson Score Interval for the accuracy of the model, which gives an interval of $[0.8891, 0.9923]$ with confidence of 99\%.

\subsection{List of Software}\label{app:los}
Document Loaders:
\begin{itemize}
    \item IBM’s Docling toolkit: \url{https://github.com/docling-project/docling}
    \item Haystack by Deepset: \url{https://haystack.deepset.ai/}  
    \item LangChain: \url{https://www.langchain.com/}
    \item LlamaIndex by LlamaIndex Inc.: \url{https://docs.llamaindex.ai/en/stable/}
    \item LLMSherpa by NLMatics: \url{https://github.com/nlmatics/llmsherpa} 
    \item NotebookLM: \url{https://notebooklm.google/}.
\end{itemize}
\subsection{RAG Implementation Details}\label{app:rag-det}
\paragraph{White-box RAG Pipelines:}These pipelines provided full access to internal components, including the retriever and language model. We tested three white-box configurations: \textbf{Llama 3.2} (3.21B parameters, Q4\_K\_M quantization—a 4-bit integer quantization with mixed-precision weight compression) with a local retriever; \textbf{Gemma 3} (27.2B parameters, Q4\_0 quantization—standard 4-bit) with a local retriever (both at temperature 1.00); and \textbf{DeepSeek R1} accessed via API, also using a local retriever. For our local (white-box) RAG pipeline implementations, we followed the guidelines in the official LangChain documentation and used a Chroma retriever with a default top $k=4$ setting. In the public repository, we report the full system prompt used with the RAG pipelines that supported user-defined prompts.

\paragraph{Black-box RAG Pipelines:} These pipelines involved using existing platforms or APIs where the internal workings of the retriever and language model were not directly accessible. We tested three black-box configurations: OpenAI GPT-4o (\texttt{gpt-4o-2024-08-06}), utilizing the assistant functionality on the OpenAI platform with default parameters (including a temperature of 1.00 and a default retrieval of 20); OpenAI o3-mini (\texttt{o3-mini-2025-01-31}), utilizing its built-in assistant functionality on the OpenAI platform with default parameters (including a temperature of 1.00 and a default retrieval of 20); and NotebookLM, a fully deployed RAG system with an online interface, based on Google's Gemini 2.0 model.

\subsection{Experiment 3: Scenarios}\label{app:exp3-details}
\subsubsection{Attack Scenarios}
We evaluate nine attacks targeting RAG pipelines using a dataset of political biographies. Each scenario tests vulnerabilities via prompt injection, obfuscation, or document-level manipulation.

\paragraph{a1-Pipeline Failure} A 169\,MB PDF embeds 97M characters out-of-bounds, causing infinite parsing or system crashes. \textit{Obfuscation}: / \quad \textit{Injection}: out-of-bound text

\paragraph{a2-Reasoning Overload} A Sudoku puzzle is injected with font poisoning; success = excessive reasoning tokens. \textit{Obfuscation}: font poisoning \quad \textit{Injection}: zero-size font, font poisoning

\paragraph{a3-Unreadable Output} Font poisoning and transparent text cause base64 gibberish replies about Draghi. \textit{Obfuscation}: font poisoning \quad \textit{Injection}: transparent text, font poisoning

\paragraph{a4-Empty Statement Response} Obfuscation hides von der Leyen’s birth info, leading to “unavailable” replies. \textit{Obfuscation}: reordering characters, zero-width characters \quad \textit{Injection}: out-of-bound text

\paragraph{a5-Vague Output} Von der Leyen’s bio altered with vague language via font poisoning. \textit{Obfuscation}: font poisoning, transparent text \quad \textit{Injection}: font poisoning

\paragraph{a6-Bias Injection} Hidden praise in Draghi’s bio biases responses. \textit{Obfuscation}: camouflage, font poisoning, metadata, zero-size font \quad \textit{Injection}: font poisoning

\paragraph{a7-Factual Distortion} Fake event (cooking show) replaces real info in Evo Morales’ bio. \textit{Obfuscation}: font poisoning \quad \textit{Injection}: font poisoning

\paragraph{a8-Outdated Knowledge} Sassoli’s death removed using homoglyphs, making the model say he is alive. \textit{Obfuscation}: homoglyph characters, zero-width characters \quad \textit{Injection}: out-of-bound text

\paragraph{a9-Sensitive Data Disclosure} Prompt injection leaks another user's PII when querying for Elena Bianchi. \textit{Obfuscation}: out-of-bound text \quad \textit{Injection}: /

%% file: main.bbl

\begin{thebibliography}{28}


\ifx \showCODEN    \undefined \def \showCODEN     #1{\unskip}     \fi
\ifx \showISBNx    \undefined \def \showISBNx     #1{\unskip}     \fi
\ifx \showISBNxiii \undefined \def \showISBNxiii  #1{\unskip}     \fi
\ifx \showISSN     \undefined \def \showISSN      #1{\unskip}     \fi
\ifx \showLCCN     \undefined \def \showLCCN      #1{\unskip}     \fi
\ifx \shownote     \undefined \def \shownote      #1{#1}          \fi
\ifx \showarticletitle \undefined \def \showarticletitle #1{#1}   \fi
\ifx \showURL      \undefined \def \showURL       {\relax}        \fi
\providecommand\bibfield[2]{#2}
\providecommand\bibinfo[2]{#2}
\providecommand\natexlab[1]{#1}
\providecommand\showeprint[2][]{arXiv:#2}

\bibitem[Auer et~al\mbox{.}(2024)]%
        {docling}
\bibfield{author}{\bibinfo{person}{Christoph Auer}, \bibinfo{person}{Maksym Lysak}, \bibinfo{person}{Ahmed Nassar}, \bibinfo{person}{Michele Dolfi}, \bibinfo{person}{Nikolaos Livathinos}, \bibinfo{person}{Panos Vagenas}, \bibinfo{person}{Cesar~Berrospi Ramis}, \bibinfo{person}{Matteo Omenetti}, \bibinfo{person}{Fabian Lindlbauer}, \bibinfo{person}{Kasper Dinkla}, \bibinfo{person}{Lokesh Mishra}, \bibinfo{person}{Yusik Kim}, \bibinfo{person}{Shubham Gupta}, \bibinfo{person}{Rafael~Teixeira de Lima}, \bibinfo{person}{Valery Weber}, \bibinfo{person}{Lucas Morin}, \bibinfo{person}{Ingmar Meijer}, \bibinfo{person}{Viktor Kuropiatnyk}, {and} \bibinfo{person}{Peter W.~J. Staar}.} \bibinfo{year}{2024}\natexlab{}.
\newblock \bibinfo{title}{Docling Technical Report}.
\newblock
\showeprint[arxiv]{2408.09869}~[cs.CL]
\urldef\tempurl%
\url{https://arxiv.org/abs/2408.09869}
\showURL{%
\tempurl}


\bibitem[Boucher et~al\mbox{.}(2023)]%
        {boucher2023boosting}
\bibfield{author}{\bibinfo{person}{Nicholas Boucher}, \bibinfo{person}{Luca Pajola}, \bibinfo{person}{Ilia Shumailov}, \bibinfo{person}{Ross Anderson}, {and} \bibinfo{person}{Mauro Conti}.} \bibinfo{year}{2023}\natexlab{}.
\newblock \showarticletitle{Boosting big brother: Attacking search engines with encodings}. In \bibinfo{booktitle}{\emph{Proceedings of the 26th International Symposium on Research in Attacks, Intrusions and Defenses}}. \bibinfo{pages}{700--713}.
\newblock


\bibitem[Boucher et~al\mbox{.}(2022)]%
        {boucher2022bad}
\bibfield{author}{\bibinfo{person}{Nicholas Boucher}, \bibinfo{person}{Ilia Shumailov}, \bibinfo{person}{Ross Anderson}, {and} \bibinfo{person}{Nicolas Papernot}.} \bibinfo{year}{2022}\natexlab{}.
\newblock \showarticletitle{Bad characters: Imperceptible nlp attacks}. In \bibinfo{booktitle}{\emph{2022 IEEE Symposium on Security and Privacy (SP)}}. IEEE, \bibinfo{pages}{1987--2004}.
\newblock


\bibitem[Brown et~al\mbox{.}(2020)]%
        {brown2020language}
\bibfield{author}{\bibinfo{person}{Tom Brown}, \bibinfo{person}{Benjamin Mann}, \bibinfo{person}{Nick Ryder}, \bibinfo{person}{Melanie Subbiah}, \bibinfo{person}{Jared~D Kaplan}, \bibinfo{person}{Prafulla Dhariwal}, \bibinfo{person}{Arvind Neelakantan}, \bibinfo{person}{Pranav Shyam}, \bibinfo{person}{Girish Sastry}, \bibinfo{person}{Amanda Askell}, {et~al\mbox{.}}} \bibinfo{year}{2020}\natexlab{}.
\newblock \showarticletitle{Language models are few-shot learners}.
\newblock \bibinfo{journal}{\emph{Advances in neural information processing systems}}  \bibinfo{volume}{33} (\bibinfo{year}{2020}), \bibinfo{pages}{1877--1901}.
\newblock


\bibitem[Carlini et~al\mbox{.}(2021)]%
        {carlini2021extracting}
\bibfield{author}{\bibinfo{person}{Nicholas Carlini}, \bibinfo{person}{Florian Tramer}, \bibinfo{person}{Eric Wallace}, \bibinfo{person}{Matthew Jagielski}, \bibinfo{person}{Ariel Herbert-Voss}, \bibinfo{person}{Katherine Lee}, \bibinfo{person}{Adam Roberts}, \bibinfo{person}{Tom Brown}, \bibinfo{person}{Dawn Song}, \bibinfo{person}{Ulfar Erlingsson}, {et~al\mbox{.}}} \bibinfo{year}{2021}\natexlab{}.
\newblock \showarticletitle{Extracting training data from large language models}. In \bibinfo{booktitle}{\emph{30th USENIX security symposium (USENIX Security 21)}}. \bibinfo{pages}{2633--2650}.
\newblock


\bibitem[Chaudhari et~al\mbox{.}(2024)]%
        {chaudhari2024phantom}
\bibfield{author}{\bibinfo{person}{Harsh Chaudhari}, \bibinfo{person}{Giorgio Severi}, \bibinfo{person}{John Abascal}, \bibinfo{person}{Matthew Jagielski}, \bibinfo{person}{Christopher~A Choquette-Choo}, \bibinfo{person}{Milad Nasr}, \bibinfo{person}{Cristina Nita-Rotaru}, {and} \bibinfo{person}{Alina Oprea}.} \bibinfo{year}{2024}\natexlab{}.
\newblock \showarticletitle{Phantom: General trigger attacks on retrieval augmented language generation}.
\newblock \bibinfo{journal}{\emph{arXiv preprint arXiv:2405.20485}} (\bibinfo{year}{2024}).
\newblock


\bibitem[Conti et~al\mbox{.}(2023)]%
        {conti2023turning}
\bibfield{author}{\bibinfo{person}{Mauro Conti}, \bibinfo{person}{Luca Pajola}, {and} \bibinfo{person}{Pier~Paolo Tricomi}.} \bibinfo{year}{2023}\natexlab{}.
\newblock \showarticletitle{Turning captchas against humanity: Captcha-based attacks in online social media}.
\newblock \bibinfo{journal}{\emph{Online Social Networks and Media}}  \bibinfo{volume}{36} (\bibinfo{year}{2023}), \bibinfo{pages}{100252}.
\newblock


\bibitem[Deng et~al\mbox{.}(2024)]%
        {pandora}
\bibfield{author}{\bibinfo{person}{Gelei Deng}, \bibinfo{person}{Yi Liu}, \bibinfo{person}{Kailong Wang}, \bibinfo{person}{Yuekang Li}, \bibinfo{person}{Tianwei Zhang}, {and} \bibinfo{person}{Yang Liu}.} \bibinfo{year}{2024}\natexlab{}.
\newblock \showarticletitle{Pandora: Jailbreak gpts by retrieval augmented generation poisoning}.
\newblock \bibinfo{journal}{\emph{arXiv preprint arXiv:2402.08416}} (\bibinfo{year}{2024}).
\newblock


\bibitem[Gr{\"o}ndahl et~al\mbox{.}(2018)]%
        {grondahl2018all}
\bibfield{author}{\bibinfo{person}{Tommi Gr{\"o}ndahl}, \bibinfo{person}{Luca Pajola}, \bibinfo{person}{Mika Juuti}, \bibinfo{person}{Mauro Conti}, {and} \bibinfo{person}{N Asokan}.} \bibinfo{year}{2018}\natexlab{}.
\newblock \showarticletitle{All you need is" love" evading hate speech detection}. In \bibinfo{booktitle}{\emph{Proceedings of the 11th ACM workshop on artificial intelligence and security}}. \bibinfo{pages}{2--12}.
\newblock


\bibitem[Hou et~al\mbox{.}(2024)]%
        {hou2024bridging}
\bibfield{author}{\bibinfo{person}{Yupeng Hou}, \bibinfo{person}{Jiacheng Li}, \bibinfo{person}{Zhankui He}, \bibinfo{person}{An Yan}, \bibinfo{person}{Xiusi Chen}, {and} \bibinfo{person}{Julian McAuley}.} \bibinfo{year}{2024}\natexlab{}.
\newblock \showarticletitle{Bridging Language and Items for Retrieval and Recommendation}.
\newblock \bibinfo{journal}{\emph{arXiv preprint arXiv:2403.03952}} (\bibinfo{year}{2024}).
\newblock


\bibitem[Huang and Chang(2022)]%
        {huang2022towards}
\bibfield{author}{\bibinfo{person}{Jie Huang} {and} \bibinfo{person}{Kevin Chen-Chuan Chang}.} \bibinfo{year}{2022}\natexlab{}.
\newblock \showarticletitle{Towards reasoning in large language models: A survey}.
\newblock \bibinfo{journal}{\emph{arXiv preprint arXiv:2212.10403}} (\bibinfo{year}{2022}).
\newblock


\bibitem[Kumar et~al\mbox{.}(2025)]%
        {kumar2025overthink}
\bibfield{author}{\bibinfo{person}{Abhinav Kumar}, \bibinfo{person}{Jaechul Roh}, \bibinfo{person}{Ali Naseh}, \bibinfo{person}{Marzena Karpinska}, \bibinfo{person}{Mohit Iyyer}, \bibinfo{person}{Amir Houmansadr}, {and} \bibinfo{person}{Eugene Bagdasarian}.} \bibinfo{year}{2025}\natexlab{}.
\newblock \showarticletitle{OverThink: Slowdown Attacks on Reasoning LLMs}.
\newblock \bibinfo{journal}{\emph{arXiv e-prints}} (\bibinfo{year}{2025}), \bibinfo{pages}{arXiv--2502}.
\newblock


\bibitem[Lazer et~al\mbox{.}(2018)]%
        {lazer2018science}
\bibfield{author}{\bibinfo{person}{David~MJ Lazer}, \bibinfo{person}{Matthew~A Baum}, \bibinfo{person}{Yochai Benkler}, \bibinfo{person}{Adam~J Berinsky}, \bibinfo{person}{Kelly~M Greenhill}, \bibinfo{person}{Filippo Menczer}, \bibinfo{person}{Miriam~J Metzger}, \bibinfo{person}{Brendan Nyhan}, \bibinfo{person}{Gordon Pennycook}, \bibinfo{person}{David Rothschild}, {et~al\mbox{.}}} \bibinfo{year}{2018}\natexlab{}.
\newblock \showarticletitle{The science of fake news}.
\newblock \bibinfo{journal}{\emph{Science}} \bibinfo{volume}{359}, \bibinfo{number}{6380} (\bibinfo{year}{2018}), \bibinfo{pages}{1094--1096}.
\newblock


\bibitem[Lewis et~al\mbox{.}(2020)]%
        {lewis2020retrieval}
\bibfield{author}{\bibinfo{person}{Patrick Lewis}, \bibinfo{person}{Ethan Perez}, \bibinfo{person}{Aleksandra Piktus}, \bibinfo{person}{Fabio Petroni}, \bibinfo{person}{Vladimir Karpukhin}, \bibinfo{person}{Naman Goyal}, \bibinfo{person}{Heinrich K{\"u}ttler}, \bibinfo{person}{Mike Lewis}, \bibinfo{person}{Wen-tau Yih}, \bibinfo{person}{Tim Rockt{\"a}schel}, {et~al\mbox{.}}} \bibinfo{year}{2020}\natexlab{}.
\newblock \showarticletitle{Retrieval-augmented generation for knowledge-intensive nlp tasks}.
\newblock \bibinfo{journal}{\emph{Advances in Neural Information Processing Systems}}  \bibinfo{volume}{33} (\bibinfo{year}{2020}), \bibinfo{pages}{9459--9474}.
\newblock


\bibitem[Livathinos et~al\mbox{.}(2025)]%
        {docling2}
\bibfield{author}{\bibinfo{person}{Nikolaos Livathinos}, \bibinfo{person}{Christoph Auer}, \bibinfo{person}{Maksym Lysak}, \bibinfo{person}{Ahmed Nassar}, \bibinfo{person}{Michele Dolfi}, \bibinfo{person}{Panos Vagenas}, \bibinfo{person}{Cesar~Berrospi Ramis}, \bibinfo{person}{Matteo Omenetti}, \bibinfo{person}{Kasper Dinkla}, \bibinfo{person}{Yusik Kim}, \bibinfo{person}{Shubham Gupta}, \bibinfo{person}{Rafael~Teixeira de Lima}, \bibinfo{person}{Valery Weber}, \bibinfo{person}{Lucas Morin}, \bibinfo{person}{Ingmar Meijer}, \bibinfo{person}{Viktor Kuropiatnyk}, {and} \bibinfo{person}{Peter W.~J. Staar}.} \bibinfo{year}{2025}\natexlab{}.
\newblock \bibinfo{title}{Docling: An Efficient Open-Source Toolkit for AI-driven Document Conversion}.
\newblock
\showeprint[arxiv]{2501.17887}~[cs.CL]
\urldef\tempurl%
\url{https://arxiv.org/abs/2501.17887}
\showURL{%
\tempurl}


\bibitem[Markwood et~al\mbox{.}(2017)]%
        {markwood2017mirage}
\bibfield{author}{\bibinfo{person}{Ian Markwood}, \bibinfo{person}{Dakun Shen}, \bibinfo{person}{Yao Liu}, {and} \bibinfo{person}{Zhuo Lu}.} \bibinfo{year}{2017}\natexlab{}.
\newblock \showarticletitle{Mirage: Content masking attack against $\{$Information-Based$\}$ online services}. In \bibinfo{booktitle}{\emph{26th USENIX Security Symposium (USENIX Security 17)}}. \bibinfo{pages}{833--847}.
\newblock


\bibitem[Mehrotra et~al\mbox{.}(2024)]%
        {mehrotra2024tree}
\bibfield{author}{\bibinfo{person}{Anay Mehrotra}, \bibinfo{person}{Manolis Zampetakis}, \bibinfo{person}{Paul Kassianik}, \bibinfo{person}{Blaine Nelson}, \bibinfo{person}{Hyrum Anderson}, \bibinfo{person}{Yaron Singer}, {and} \bibinfo{person}{Amin Karbasi}.} \bibinfo{year}{2024}\natexlab{}.
\newblock \showarticletitle{Tree of attacks: Jailbreaking black-box llms automatically}.
\newblock \bibinfo{journal}{\emph{Advances in Neural Information Processing Systems}}  \bibinfo{volume}{37} (\bibinfo{year}{2024}), \bibinfo{pages}{61065--61105}.
\newblock


\bibitem[Pajola and Conti(2021)]%
        {pajola2021fall}
\bibfield{author}{\bibinfo{person}{Luca Pajola} {and} \bibinfo{person}{Mauro Conti}.} \bibinfo{year}{2021}\natexlab{}.
\newblock \showarticletitle{Fall of Giants: How popular text-based MLaaS fall against a simple evasion attack}. In \bibinfo{booktitle}{\emph{2021 IEEE European Symposium on Security and Privacy (EuroS\&P)}}. IEEE, \bibinfo{pages}{198--211}.
\newblock


\bibitem[Perez and Ribeiro(2022)]%
        {perez2022ignore}
\bibfield{author}{\bibinfo{person}{F{\'a}bio Perez} {and} \bibinfo{person}{Ian Ribeiro}.} \bibinfo{year}{2022}\natexlab{}.
\newblock \showarticletitle{Ignore previous prompt: Attack techniques for language models}.
\newblock \bibinfo{journal}{\emph{arXiv preprint arXiv:2211.09527}} (\bibinfo{year}{2022}).
\newblock


\bibitem[Roller et~al\mbox{.}(2021)]%
        {roller2020recipes}
\bibfield{author}{\bibinfo{person}{Stephen Roller}, \bibinfo{person}{Emily Dinan}, \bibinfo{person}{Naman Goyal}, \bibinfo{person}{Da Ju}, \bibinfo{person}{Mary Williamson}, \bibinfo{person}{Yinhan Liu}, \bibinfo{person}{Jing Xu}, \bibinfo{person}{Myle Ott}, \bibinfo{person}{Kurt Shuster}, \bibinfo{person}{Eric~M Smith}, {et~al\mbox{.}}} \bibinfo{year}{2021}\natexlab{}.
\newblock \showarticletitle{Recipes for building an open-domain chatbot}.
\newblock \bibinfo{journal}{\emph{Proceedings of the 16th Conference of the European Chapter of the Association for Computational Linguistics: Main Volume}} (\bibinfo{year}{2021}).
\newblock


\bibitem[Wei et~al\mbox{.}(2023)]%
        {wei2023jailbroken}
\bibfield{author}{\bibinfo{person}{Alexander Wei}, \bibinfo{person}{Nika Haghtalab}, {and} \bibinfo{person}{Jacob Steinhardt}.} \bibinfo{year}{2023}\natexlab{}.
\newblock \showarticletitle{Jailbroken: How does llm safety training fail?}
\newblock \bibinfo{journal}{\emph{Advances in Neural Information Processing Systems}}  \bibinfo{volume}{36} (\bibinfo{year}{2023}), \bibinfo{pages}{80079--80110}.
\newblock


\bibitem[Wei et~al\mbox{.}(2022)]%
        {wei2022chain}
\bibfield{author}{\bibinfo{person}{Jason Wei}, \bibinfo{person}{Xuezhi Wang}, \bibinfo{person}{Dale Schuurmans}, \bibinfo{person}{Maarten Bosma}, \bibinfo{person}{Fei Xia}, \bibinfo{person}{Ed Chi}, \bibinfo{person}{Quoc~V Le}, \bibinfo{person}{Denny Zhou}, {et~al\mbox{.}}} \bibinfo{year}{2022}\natexlab{}.
\newblock \showarticletitle{Chain-of-thought prompting elicits reasoning in large language models}.
\newblock \bibinfo{journal}{\emph{Advances in neural information processing systems}}  \bibinfo{volume}{35} (\bibinfo{year}{2022}), \bibinfo{pages}{24824--24837}.
\newblock


\bibitem[Weidinger et~al\mbox{.}(2022)]%
        {weidinger2022taxonomy}
\bibfield{author}{\bibinfo{person}{Laura Weidinger}, \bibinfo{person}{Jonathan Uesato}, \bibinfo{person}{Maribeth Rauh}, \bibinfo{person}{Conor Griffin}, \bibinfo{person}{Po-Sen Huang}, \bibinfo{person}{John Mellor}, \bibinfo{person}{Amelia Glaese}, \bibinfo{person}{Myra Cheng}, \bibinfo{person}{Borja Balle}, \bibinfo{person}{Atoosa Kasirzadeh}, {et~al\mbox{.}}} \bibinfo{year}{2022}\natexlab{}.
\newblock \showarticletitle{Taxonomy of risks posed by language models}. In \bibinfo{booktitle}{\emph{Proceedings of the 2022 ACM conference on fairness, accountability, and transparency}}. \bibinfo{pages}{214--229}.
\newblock


\bibitem[Xiao et~al\mbox{.}(2018)]%
        {xiao2018security}
\bibfield{author}{\bibinfo{person}{Qixue Xiao}, \bibinfo{person}{Kang Li}, \bibinfo{person}{Deyue Zhang}, {and} \bibinfo{person}{Weilin Xu}.} \bibinfo{year}{2018}\natexlab{}.
\newblock \showarticletitle{Security risks in deep learning implementations}. In \bibinfo{booktitle}{\emph{2018 IEEE Security and privacy workshops (SPW)}}. IEEE, \bibinfo{pages}{123--128}.
\newblock


\bibitem[Xue et~al\mbox{.}(2024)]%
        {badrag}
\bibfield{author}{\bibinfo{person}{Jiaqi Xue}, \bibinfo{person}{Mengxin Zheng}, \bibinfo{person}{Yebowen Hu}, \bibinfo{person}{Fei Liu}, \bibinfo{person}{Xun Chen}, {and} \bibinfo{person}{Qian Lou}.} \bibinfo{year}{2024}\natexlab{}.
\newblock \showarticletitle{Badrag: Identifying vulnerabilities in retrieval augmented generation of large language models}.
\newblock \bibinfo{journal}{\emph{arXiv preprint arXiv:2406.00083}} (\bibinfo{year}{2024}).
\newblock


\bibitem[Yao et~al\mbox{.}(2024)]%
        {yao2024survey}
\bibfield{author}{\bibinfo{person}{Yifan Yao}, \bibinfo{person}{Jinhao Duan}, \bibinfo{person}{Kaidi Xu}, \bibinfo{person}{Yuanfang Cai}, \bibinfo{person}{Zhibo Sun}, {and} \bibinfo{person}{Yue Zhang}.} \bibinfo{year}{2024}\natexlab{}.
\newblock \showarticletitle{A survey on large language model (llm) security and privacy: The good, the bad, and the ugly}.
\newblock \bibinfo{journal}{\emph{High-Confidence Computing}} (\bibinfo{year}{2024}), \bibinfo{pages}{100211}.
\newblock


\bibitem[Zheng et~al\mbox{.}(2023)]%
        {zheng2023judging}
\bibfield{author}{\bibinfo{person}{Lianmin Zheng}, \bibinfo{person}{Wei-Lin Chiang}, \bibinfo{person}{Ying Sheng}, \bibinfo{person}{Siyuan Zhuang}, \bibinfo{person}{Zhanghao Wu}, \bibinfo{person}{Yonghao Zhuang}, \bibinfo{person}{Zi Lin}, \bibinfo{person}{Zhuohan Li}, \bibinfo{person}{Dacheng Li}, \bibinfo{person}{Eric Xing}, {et~al\mbox{.}}} \bibinfo{year}{2023}\natexlab{}.
\newblock \showarticletitle{Judging llm-as-a-judge with mt-bench and chatbot arena}.
\newblock \bibinfo{journal}{\emph{Advances in Neural Information Processing Systems}}  \bibinfo{volume}{36} (\bibinfo{year}{2023}), \bibinfo{pages}{46595--46623}.
\newblock


\bibitem[Zou et~al\mbox{.}(2024)]%
        {poisonedrag}
\bibfield{author}{\bibinfo{person}{Wei Zou}, \bibinfo{person}{Runpeng Geng}, \bibinfo{person}{Binghui Wang}, {and} \bibinfo{person}{Jinyuan Jia}.} \bibinfo{year}{2024}\natexlab{}.
\newblock \showarticletitle{Poisonedrag: Knowledge corruption attacks to retrieval-augmented generation of large language models}.
\newblock \bibinfo{journal}{\emph{arXiv preprint arXiv:2402.07867}} (\bibinfo{year}{2024}).
\newblock


\end{thebibliography}
